\documentclass[11pt,aps,prd,preprint,preprintnumbers,showpacs,superscriptaddress,floatfix]{revtex4-1}
\usepackage{graphicx,}
\usepackage{graphicx,dcolumn,bm,epsfig,amsmath,amssymb,}
\usepackage{epsfig}
\usepackage{subfigure}
\usepackage{booktabs}
\usepackage{multirow}
\usepackage{float}
\usepackage{dcolumn}
\usepackage{bm}
\usepackage{amsmath}
\usepackage{amssymb}
\usepackage{bbm}
\usepackage{multirow}
\usepackage{setspace}
\usepackage{xcolor}
\usepackage[colorlinks,
            linkbordercolor={0 0 0},
            pdfborder={0 0 0},
            linkcolor=blue,
            citecolor=blue,
            urlcolor=blue,
            breaklinks=true]{hyperref}

\usepackage{tikz}
\usetikzlibrary{positioning,decorations.pathmorphing,decorations.markings,arrows}
\tikzset{
    vector/.style={decorate, decoration={snake}, draw},
    provector/.style={decorate, decoration={snake,amplitude=2.5pt}, draw},
    antivector/.style={decorate, decoration={snake,amplitude=-2.5pt}, draw},
    fermion/.style={draw=black,
      postaction={decorate},decoration={markings,mark=at position .55
        with {\arrow[draw=black]{>}}}},
    fermionbar/.style={draw=black, postaction={decorate},
                       decoration={markings,mark=at position .55 with {\arrow[draw=black]{<}}}},
    fermionnoarrow/.style={draw=black},
    gluon/.style={decorate, draw=black,decoration={coil,amplitude=4pt, segment length=6pt}},
    scalar/.style={dashed,draw=black,
      postaction={decorate},decoration={markings,mark=at position .55
        with {\arrow[draw=black]{>}}}},
    scalarbar/.style={dashed,draw=black,
      postaction={decorate},decoration={markings,mark=at position .55
        with {\arrow[draw=black]{<}}}},
    scalarnoarrow/.style={dashed,draw=black},
    electron/.style={draw=black,
      postaction={decorate},decoration={markings,mark=at position .55
        with {\arrow[draw=black]{>}}}},
    bigvector/.style={decorate, decoration={snake,amplitude=4pt}, draw},
}

\newlength{\x}
\newlength{\y}
\newlength{\z}
\phantomsection

\begin{document}
\preprint{IMSc/2022/04}

\title {Confinement-deconfinement transition  in
$SU(3)$-Higgs theory}

\author{Sanatan Digal}
\email{digal@imsc.res.in}              
\affiliation{The Institute of Mathematical Sciences, Chennai 600113, India}
\affiliation{Homi Bhabha National Institute, Training School Complex,
Anushakti Nagar, Mumbai 400094, India}

\author{Vinod Mamale}
\email{mvinod@imsc.res.in}
\affiliation{The Institute of Mathematical Sciences, Chennai 600113, India}
\affiliation{Homi Bhabha National Institute, Training School Complex,
Anushakti Nagar, Mumbai 400094, India}

\author{Sabiar Shaikh}
\email{sabiarshaikh@imsc.res.in}
\affiliation{The Institute of Mathematical Sciences, Chennai 600113, India}
\affiliation{Homi Bhabha National Institute, Training School Complex,
Anushakti Nagar, Mumbai 400094, India}
\affiliation{School of Physical Sciences, National Institute of Science Education and Research, Jatni 752050, India}

\begin{abstract}

We study lattice cutoff effects on the confinement-deconfinement transition and the $Z_3$ symmetry in $SU(3)$-Higgs theory in $3+1$ dimensions. The Higgs in this study is a complex triplet with vanishing bare mass and quartic coupling. The lattice cutoff is regulated by varying the number of temporal lattice sites, $N_\tau$. Our results show that the nature of the confinement-deconfinement transition depends on $N_\tau$. For $N_\tau=2$ the transition is found to be the end point of a first-order transition and is first order for $N_\tau \ge 3$. The distributions of the Polyakov loop and other observables, sensitive to the $Z_3$ symmetry, show that the strength of $Z_3$ explicit breaking decreases with $N_\tau$.  Up to $T\simeq 2T_c$, the free energy difference between $Z_3$ states decreases with $N_\tau$, suggesting the realization of $Z_3$ symmetry in the continuum limit.

\end{abstract}
\pacs{}

\maketitle

\section{Introduction}

Studies of the confinement-deconfinement (CD) transition in quantum chromodynamics, electroweak theory, etc., are key to understanding matter at extreme temperatures. These studies are also important for the phase diagram of these theories. It is well known that the transition, from a confined to a deconfined phase, is primarily driven by the non-Abelian gauge fields. Its nature depends on the gauge group $SU(N)$, couplings, and masses of the matter fields in the theory~\cite{Nakamura:1984uz,Fukugita:1986rr,Kogut:1982rt,Karsch:2001nf}. In the pure gauge limit, the 
confined and deconfined phases are characterized by the free energy of an isolated static charge. At low temperatures, the free energy diverges, which leads to confinement. In the string model of confinement, color singlet pairs of static charges are connected by a string of nonzero tension. This is backed by the first principle lattice gauge theory calculations that show the free energy of the pairs rising linearly with separation ($r$) between them~\cite{Polyakov:1978vu,Susskind:1979up,Green:1982hu}. Thermal fluctuations at high temperatures melt the string, which leads to the liberation of the static charges and the onset of deconfinement~\cite{Pisarski:1982cn}. In the Euclidean formulation of thermodynamics, the free energy of a static charge in units of temperature is given by the negative logarithm of the Polyakov loop thermal average~\cite{Polyakov:1978vu,Susskind:1979up,McLerran:1981pb,McLerran:1980pk,Gavai:1985nz}. As this average vanishes in the confined phase and acquires a nonzero value in the deconfined phase, it acts as an order parameter for the CD transition~\cite{Kuti:1980gh,McLerran:1980pk,Weiss:1980rj,Svetitsky:1982gs}.  Furthermore, the Polyakov loop transforms like a $Z_N$ spin under gauge transformations, that are twisted along the temporal direction by $Z_N$ phases~\cite{Svetitsky:1982gs,McLerran:1981pb,Svetitsky:1985ye}. Since the Polyakov loop acquires a nonzero average in the deconfined phase, the $Z_N$ symmetry is spontaneously broken, which subsequently leads to $N$ degenerate $Z_N$ states~\cite{Yaffe:1982qf,Celik:1983wz,Weiss:1980rj,Svetitsky:1985ye}.

In the presence of dynamical matter fields in the fundamental representation, the string connecting the static singlet pair breaks due to excitations of dynamical charges. The matter fields, after a twisted/$Z_N$ gauge transformation, do not satisfy necessary temporal boundary conditions~\cite{Weiss:1981ev,Belyaev:1991np,Green:1983sd,Biswal:2016xyq,Biswal:2015rul}. Thus, the transformed matter fields cannot be part of the path integral of the partition function. Nevertheless, two gauge field configurations belonging to different $Z_N$ sectors of the Polyakov loop contribute to the partition function. It is clear that the two contributions will not be the same, as only the gauge fields can be rotated by $Z_N$
gauge transformations. This suggests that the $Z_N$ symmetry is explicitly broken. But determining the strength or the extent of the explicit breaking requires integrating out the matter fields. Note that this situation is different from the explicit breaking in spin models due to the external field, which
is not a dynamical field but a constant parameter.

Studies of spin systems show that, with the increase in strength of the explicit breaking, phase transitions soften. A strong first-order phase transition turns into a crossover for large enough explicit breaking. So, it is expected that $Z_N$ explicit breaking will cause softening of the CD transition~\cite{Green:1983sd,Heller:1984eq,Kogut:1985xd,Hasenfratz:1983ce,Heller:1985wc,Gavai:1985nz}.  Also, the Polyakov loop average is expected to be nonzero even in the confined phase. There are many studies on the effect of dynamical matter fields on the $Z_N$ symmetry~\cite{Green:1983sd,Weiss:1981ev,Ignatius:1991nk,Belyaev:1991np,Dixit:1991et,Biswal:2019xju,Fradkin:1978dv,Deka:2010bc}. In lattice gauge theories in the strong coupling limit, mean-field calculations show that a decrease in quark masses increases the explicit breaking~\cite{Green:1983sd,Biswal:2018ilm}. Perturbative loop calculations also find that, with a decrease in the mass of dynamical fields, the explicit breaking increases ~\cite{Weiss:1981ev,Biswal:2018ilm,Guo:2018scp,Gross:1980br}. Furthermore, the free energy difference between the different $Z_N$ states increases with temperature.

The $Z_N$ breaking due to dynamical matter fields near the CD transition is studied mostly in nonperturbative lattice simulations. Early lattice studies of the CD transition in $SU(2)$ with dynamical quarks showed a sharp crossover~\cite{Satz:1985js}. In $SU(3)$ gauge theory with dynamical quarks, a decrease in quark masses leads to an increase in explicit breaking in the heavy-quark region~\cite{Karsch:2001nf}.  For small enough masses, the explicit breaking is so large that the CD transition becomes a crossover. In $SU(2)$-Higgs theory~\cite{Damgaard:1986jg}, the CD transition was found to be sharper for a smaller cutoff. Recent studies of $Z_2$ symmetry in $SU(2)$-Higgs theory show that the explicit breaking decreases drastically in the Higgs symmetric side of the phase diagram~\cite{Biswal:2016xyq}. This suggests that the Higgs condensate could be playing the role of the symmetry-breaking field. In these studies, the lattices used had only a few temporal lattice sites($N_\tau \le 4$). A detailed study of cutoff effects was done for vanishing bare Higgs mass ($m_H$) and quartic coupling ($\lambda$) in Ref.~\cite{Biswal:2016xyq}. It was observed that the CD transition becomes sharper with a smaller lattice cutoff, i.e., a larger $N_\tau$. Furthermore, finite-size scaling was observed near the critical point, for $N_\tau\ge 8$. The distributions of the Polyakov loop near the transition region exhibited $Z_2$ symmetry, within statistical errors, suggesting vanishingly small explicit breaking in the continuum limit, i.e., for $N_\tau \to \infty$~\cite{Biswal:2016xyq}. Note that strong coupling, as well as perturbative calculations, suggested maximal explicit $Z_2$ breaking for $m_H=0$ and $\lambda=0$. We mention here that $Z_N$ symmetry has been observed in one-dimensional gauged Higgs chains in the continuum limit~\cite{Biswal:2021mhp}. One-dimensional $Z_2$-Higgs theory also exhibits the $Z_2$ symmetry in the thermodynamic limit~\cite{Biswal:2021fde}.

It is important to explore the $Z_N$ symmetry in the continuum limit for higher $N$.  In the present work, we extend the previous work~\cite{Biswal:2016xyq} to $SU(3)$-Higgs theory. We consider a simple mass term in the Higgs potential, ignoring the 
quartic or higher order interactions. This ensures that the system is in 
the Higgs symmetric phase in the vicinity of the CD transition, where we 
intend to study the $Z_3$ explicit symmetry breaking. The higher-order 
interaction terms in the Higgs potential, such as the quartic interaction 
term ($\lambda\Phi^4$), have been used to study the explicit breaking of 
$Z_2$ symmetry on the phase diagram of  $SU(2)$-Higgs theory. This study 
found that the explicit breaking of $Z_2$ symmetry is large in the Higgs 
broken phase and gradually decreases on approach toward the Higgs transition 
point on the phase diagram. The explicit breaking is found to be vanishingly 
small in the Higgs symmetric part of the phase diagram, in the continuum limit.
We plan to work on the explicit breaking of $Z_3$ on the phase diagram of 
$SU(3)$-Higgs theory in the future. The perturbative one-loop calculations, 
for $\lambda=0$, find that the explicit breaking is maximal for vanishing bare 
Higgs mass and monotonically decreasing to a vanishingly small value in the 
quenched limit. Lattice results for a given cutoff, in the nonperturbative 
regime, also show that the explicit breaking decreases with $m_H$; thus, we 
consider the case of maximum possible explicit breaking, i.e., $m_H=0$, 
in the present study.
 As in the case of $SU(2)$-Higgs, the CD transition is found to depend on the lattice cutoff. The distributions of the Polyakov loop show that the strength of explicit breaking decreases with $N_\tau$. With the decrease in explicit breaking, the CD transition becomes stronger. The CD transition is found to be an end point of a first-order phase transition for $N_\tau=2$ and is a first-order transition for $N_\tau \ge 3$. We also compare physical observables between the different $Z_3$ states in the deconfined phase up to $T\simeq 2T_c$, which suggests that the free energy difference between them is vanishingly small in the continuum limit.\\
  This paper is organized as follows. In Sec. II, we discuss the $Z_N$ symmetry in the presence of fundamental Higgs fields. This is followed by numerical simulations of CD transition and the $Z_3$ symmetry in pure $SU(3)$ gauge theory and in $SU(3)$-Higgs theory in Sec. III. In Sec. IV, discussions and conclusions are presented.

\section{$Z_N$ SYMMETRY IN THE PRESENCE OF FUNDAMENTAL HIGGS FIELDS}
\noindent 
The path-integral form of the partition function, ${\cal Z}$, for pure $SU(N)$ gauge theory at finite temperature  is given by
\begin{equation}
{\cal Z}=\int [DA_\mu]e^{-S_E[A_\mu]}.
\label{prt1}
\end{equation}
$A_\mu = T^a A^a_\mu$, where $T^a, ~a=1,2,...,N^2-1$, are generators of $SU(N)$.  In terms of the gauge fields $A^a_\mu$, the non-Abelian field strengths $F^a_{\mu\nu}$ are written as $F^a_{\mu\nu} = \partial_\mu A^a_\nu-\partial_\nu A^a_\mu+igf^{abc}A^b_\mu A^c_\nu$.
The Euclidean action $S_E[A]$ is given by
\begin{equation}
S_E[A] = \int_V d^3x \int_0^{\beta} d\tau \Bigg[{1\over 2} {\rm Tr} [F_{\mu\nu}(\vec{x},\tau)F_{\mu\nu}(\vec{x},\tau)]\Bigg].
\label{pact}
\end{equation}
Here, $\beta$ is the inverse of temperature, i.e., $\beta={1 / T}$. The integration in Eq.~(\ref{prt1}) is carried out over gauge fields that are periodic along the temporal direction, i.e., $A_\mu(\vec{x},\tau=0)=A_\mu(\vec{x},\tau=\beta)$. 
The action Eq.~(\ref{pact}) is invariant under the following  gauge transformation of gauge fields:
\begin{equation}
A_\mu(\vec{x},\tau) \rightarrow V(\vec{x},\tau)A_\mu(\vec{x},\tau)V^{-1}(\vec{x},\tau)-{i \over g} V(\vec{x},\tau)\partial_\mu V^{-1}(\vec{x},\tau)
\label{gtr}
\end{equation}
where $V(\vec{x},\tau) \in SU(N)$. The periodicity of the gauge transformed fields is preserved, even if $V(\vec{x},\tau)$ is not periodic in $\tau$ but satisfies
\begin{equation}
 V(\vec{x},\tau=0)=zV(\vec{x},\tau=\beta),~{\rm with}~z\in Z_N\subset SU(N).
 \label{zngtr}
 \end{equation}
 Here, $z =\mathbbm{1} \exp ({2\pi i n \over N})$ with $n=0,1,2,...,N-1$. The term $Z_N$ symmetry refers to the fact that all allowed gauge transformations of the Euclidean gauge action are classified by center $Z_N$ of the gauge group $SU(N)$. Under these gauge transformations,  the Polyakov loop
\begin{equation}
L(\vec{x})={1 \over N} {\rm Tr} \Big[P \Big\{ \exp\Big(-ig\int_0^{\beta}  A_0(\vec{x},\tau)d\tau\Big)\Big\}\Big]
\end{equation}
transforms as $L \rightarrow zL$. This transformation of the Polyakov loop is similar to that of magnetization under $Z_2$ transformation in the Ising model~\cite{Svetitsky:1985ye,Svetitsky:1982gs}. As mentioned previously, the thermal average of the Polyakov loop vanishes in the confined phase. In the deconfined phase, the Polyakov loop acquires a nonzero thermal average value, which leads to the spontaneous breaking of the $Z_N$ symmetry. As a result, there are $N$ degenerate states in the deconfined phase characterized by the elements of $Z_N$. \\ 
In the presence of the Higgs field $\Phi$ in the fundamental representation, the Euclidean $SU(N)$-Higgs action is given by
\begin{equation}
S_E[A,\Phi] = S_E[A] + \int_V d^3 x \int _0^{\beta}d\tau  \Bigg[{1\over 2} (D_{\mu}\Phi)^\dagger (D_{\mu}\Phi)+\frac{m_H^2}{2} \Phi^\dagger \Phi +\frac{\lambda}{4!}(\Phi^\dagger \Phi)^2 \Bigg]
\label{tac}
\end{equation}
Here, the covariant derivative $D_{\mu}\Phi = \partial_\mu \Phi + ig A_\mu \Phi$. $m_H$ and $\lambda$ are the mass and quartic coupling of the
Higgs field, respectively.  The total partition function of this theory at finite temperature is given by
\begin{equation}
Z=\int [DA][D\Phi]e^{-S_E[A,\Phi]}.
\end{equation}
$\Phi$ satisfies periodic boundary condition in the temporal direction, i.e.,
\begin{equation}
\Phi(\vec{x},0)=\Phi(\vec{x},\beta).
\end{equation}
Under the $SU(N)$ gauge transformation, the $\Phi$ field transforms as
\begin{equation}
\Phi(\vec{x},\tau) \rightarrow \Phi^\prime(\vec{x},\tau)=V(\vec{x},\tau) \Phi(\vec{x},\tau).
\end{equation}
A twisted $Z_N$ gauge transformation, with $V(\vec{x},\tau=0)=zV(\vec{x},\tau=\beta)$ and $z\neq \mathbbm{1}$, would lead to $\Phi^\prime$ with
\begin{equation}
\Phi '(\vec{x},0)=z \Phi '(\vec{x},\beta).
\end{equation}
As $\Phi^\prime$ is not periodic, it cannot be part of the path integral of the partition function. Therefore, gauge transformations for which $z\neq \mathbbm{1}$ are not a symmetry of the action [Eq.~(\ref{tac})]. But gauge fields that are related by gauge transformations [Eq.~(\ref{zngtr})] can both
contribute to the partition function. These contributions will not be equal, as the twisted gauge transformations cannot act on the Higgs. One can
show that the difference is due to only one term in Eq.~(\ref{tac}), i.e.,
\begin{equation}
\int_V d^3 x \int _0^{\beta}d\tau  \Bigg[{1\over 2} (D_{0}\Phi)^\dagger (D_{0}\Phi)\Bigg],
\end{equation}
involving a temporal covariant derivative. Note that gauge transformations [Eq.~(\ref{zngtr})] can be written
as
\begin{equation}
V({\bf x},\tau)=V_a(\tau)V_p({\bf x},\tau),~~~V_a(\tau=0)=zV_a(\tau=\beta),~~~V_p(\tau=0)=V_p(\tau=\beta).
\end{equation}
To see the effect of $Z_N$ gauge transformations, one needs to consider only $V_a(\tau)$. Suppose
\begin{equation}
V_a(\tau=0)=\mathbbm{1} ~~~{\rm and}~~~z=\mathbbm{1}e^{(2\pi i q/N)},
\end{equation}
with $q=0,1,...,N-1$. This transformation is gauge equivalent to $V_a(\tau)={\rm exp}[i\alpha(\tau)]$, with $\alpha(\tau)=0$ for $\tau < \beta$ and $\alpha(\beta)=2\pi q/N$. This will affect only the terms in which temporal gauge fields are involved, i.e., $|D_0\Phi|^2$. So, at the leading order, the explicit breaking of $Z_N$ arises due to temporal gradient terms. 

\par To compute the strength of $Z_N$ explicit breaking, the Higgs field must be integrated out. It is possible to achieve this in simplified models, e.g., one-dimensional gauged Higgs chain with $\lambda=0$. In this case, the explicit breaking becomes vanishingly small in the continuum limit~\cite{Biswal:2021mhp}. In the Higgs symmetric phase of $Z_2$-Higgs theory,  the entropy contribution to the partition function is $Z_2$ invariant for large $N_\tau$. As the entropy dominates the Boltzmann factor in the thermodynamic limit, the $Z_2$ symmetry is realized~\cite{Biswal:2021fde}. For a $3+1$ dimensional model, it is not possible to integrate out the Higgs
field exactly. In this work, we consider simulating the partition function [Eq.~(\ref{prt1})] using Monte Carlo techniques. In the following, we describe our simulations of the partition function and results.
\section{Monte Carlo simulations of $SU(3)$-Higgs theory}
To carry out the Monte Carlo (MC) simulation, the $3+1$ dimensional Euclidean space $L^3\times\beta$ is discretized as a lattice with $N_s^3 \times N_\tau$ points. In terms of the lattice constant $a$, $N_s=(L/a)$ and $N_\tau=(\beta/a)$. The lattice sites are denoted by $n = (n_1, n_2, n_3, n_4)$ with $1 \leq n_1, n_2, n_3 \leq N_s$ and $1 \leq n_4 \leq N_\tau$. The Higgs field $\Phi_n$ lives at the site $n$, and the gauge link $U_{n,\mu}=e^{igaA_\mu({ n})}$ is between the sites ${n}$ and ${ n}+\hat\mu$. The discretized lattice action for $\lambda=0$ and $m_H=0$ takes the following form~\cite{Biswal:2016xyq,Kajantie:1995kf}:
\begin{equation}
S = \beta_g\sum_P {\rm Tr}\left(1-{U_P + U^\dag_P \over 2 }\right) - \kappa\sum_{{n},\mu} 
{\rm Re}\left(\Phi^\dag_{{ n}+\mu}U_{n,\mu}\Phi_{n}\right)+ {1 \over 2}\sum_{{ n}}(\Phi^\dag_{ n}\Phi_{ n}),
\label{lac}
\end{equation}
where the hopping parameter $\kappa=1/8$.  $\beta_g = \frac{2N}{g^2}$, where $g$ is the gauge coupling constant. The plaquette $U_P$ is the path-ordered product of links $U_{n,\mu}$ along an elementary square, i.e.,
\begin{equation}
U_P=U_{n,\mu} U_{n+\mu,\nu}U^\dag_{n+\nu,\mu}U^\dag_{n,\nu}.
\end{equation}
In the action Eq.~(\ref{lac}),  the second term corresponds to the gauge-Higgs interaction. The average of the Polyakov loop ($L$) is
given by
\begin{equation}
L = {1 \over N_s^3}\sum_{\vec{n}} L(\vec{n}),~~~~~L(\vec{n}) =\frac{1}{N} {\rm Tr} \displaystyle\prod_{n_4=1}^{N_\tau} U_{{(\vec{n},n_4)},\hat{4}}.
\end{equation}
Here, $\vec{n} \equiv n_1, n_2, n_3$ are the spatial and $n_4$ is the temporal coordinates.
\par In the Monte Carlo simulations, an initial configuration of $\{U_{n,\mu},\Phi_n\}$ is updated according to the probability distribution, Exp($-S$). To update a given link $U_{n,\mu}$, the rest of the fields coupled to it are treated as a heat bath. A new choice for the link is generated using the standard heat-bath method~\cite{Cabibbo:1982zn,Kennedy:1985nu}. In the case of components of $\Phi_n$, the new values are obtained from a Gaussian distribution, whose peak is determined by $\kappa$ and nearest-neighbor fields. This procedure is repeated sequentially for all the links and site variables, which we call a sweep. Since a new configuration is generated from an old one, the two are correlated. Based on the autocorrelation
of the Polyakov loop, many sweeps are carried out before a configuration is considered for calculating physical observables. The observables computed are the average of the magnitude of the Polyakov loop ($|L|$) and distributions $H(|L|),H({\rm Arg}(L))$, the gauge-Higgs interaction term, $S_K={\rm Re}\sum_{n,\mu}\left(\Phi^\dag_{{ n}+\mu}U_{n,\mu}\Phi_{n}\right)$, and the plaquette ($S_g=\sum_p U_p$). The simulations were carried out for several values of $N_\tau=2,3,4,8$, to study the $N_\tau$ dependence.  We set $N_s \ge 4N_\tau$ for all the simulations. Pure $SU(3)$ 
simulations were carried out to observe the effects of the Higgs field. In the following, we present our results.
\subsection{The CD transition vs  $N_\tau$}
\begin{figure}[h]
  \centering
    \begin{minipage}[b]{0.4\textwidth}
    \includegraphics[width=\textwidth]{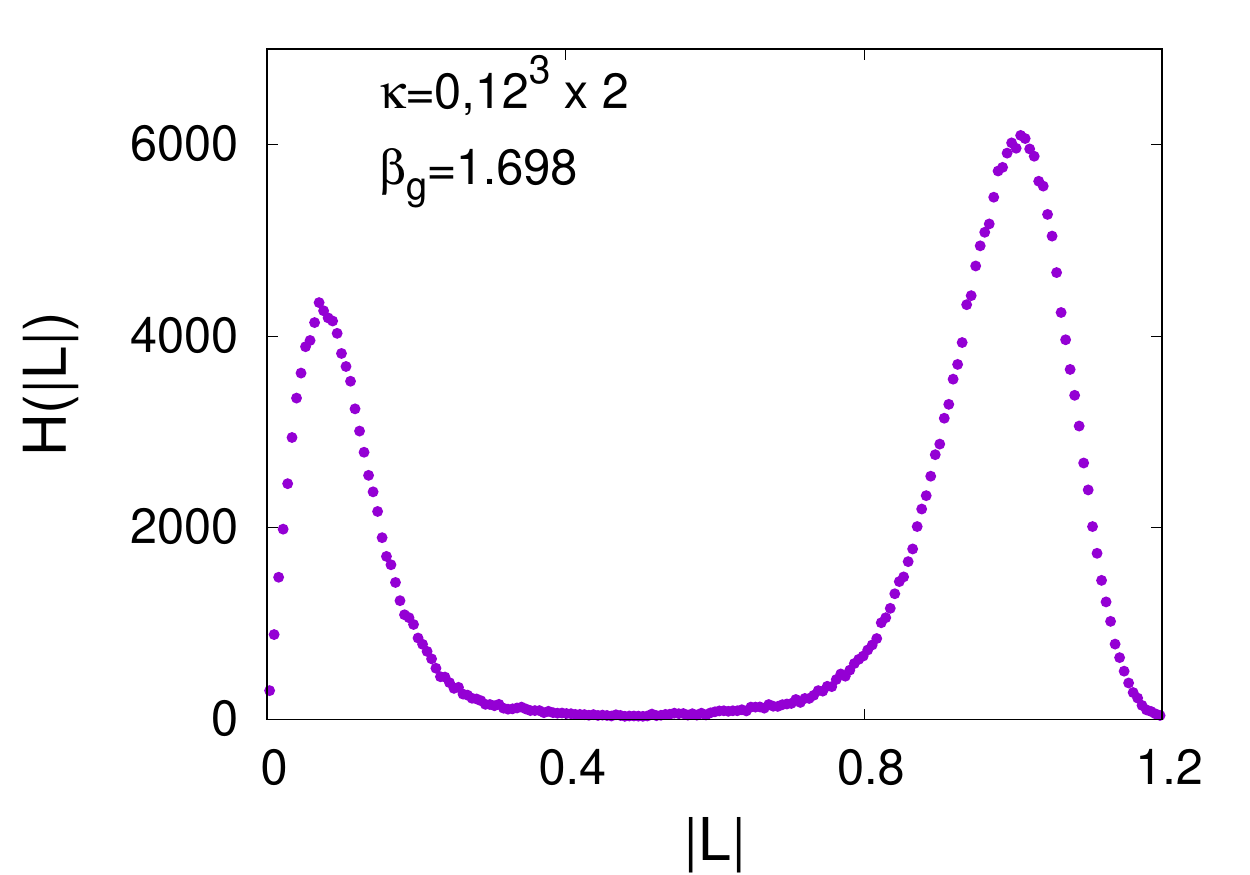}
    \caption{$H(|L|)$ for $N_\tau=2$ at $\beta_g=1.698$ and $\kappa=0$.}
    \label{plhst_nt2}
  \end{minipage}
  \hfill
  \begin{minipage}[b]{0.4\textwidth}
    \includegraphics[width=\textwidth]{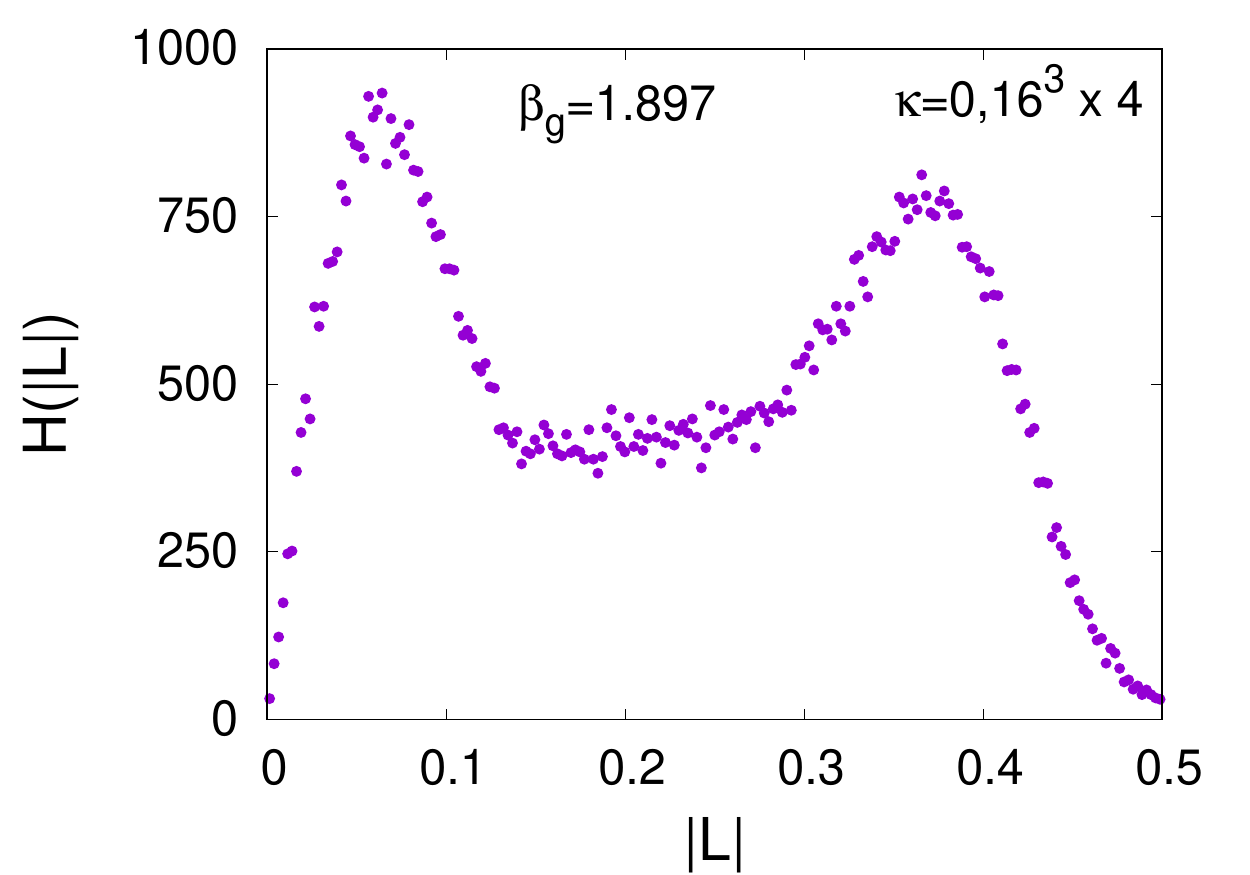}
    \caption{$H(|L|)$ for $N_\tau=4$ at $\beta_g=1.897$ and $\kappa=0$.}
    \label{plhst_nt4}
  \end{minipage}
  \end{figure}
\begin{figure}[h]
  \centering
    \begin{minipage}[b]{0.48\textwidth}
    \includegraphics[width=\textwidth]{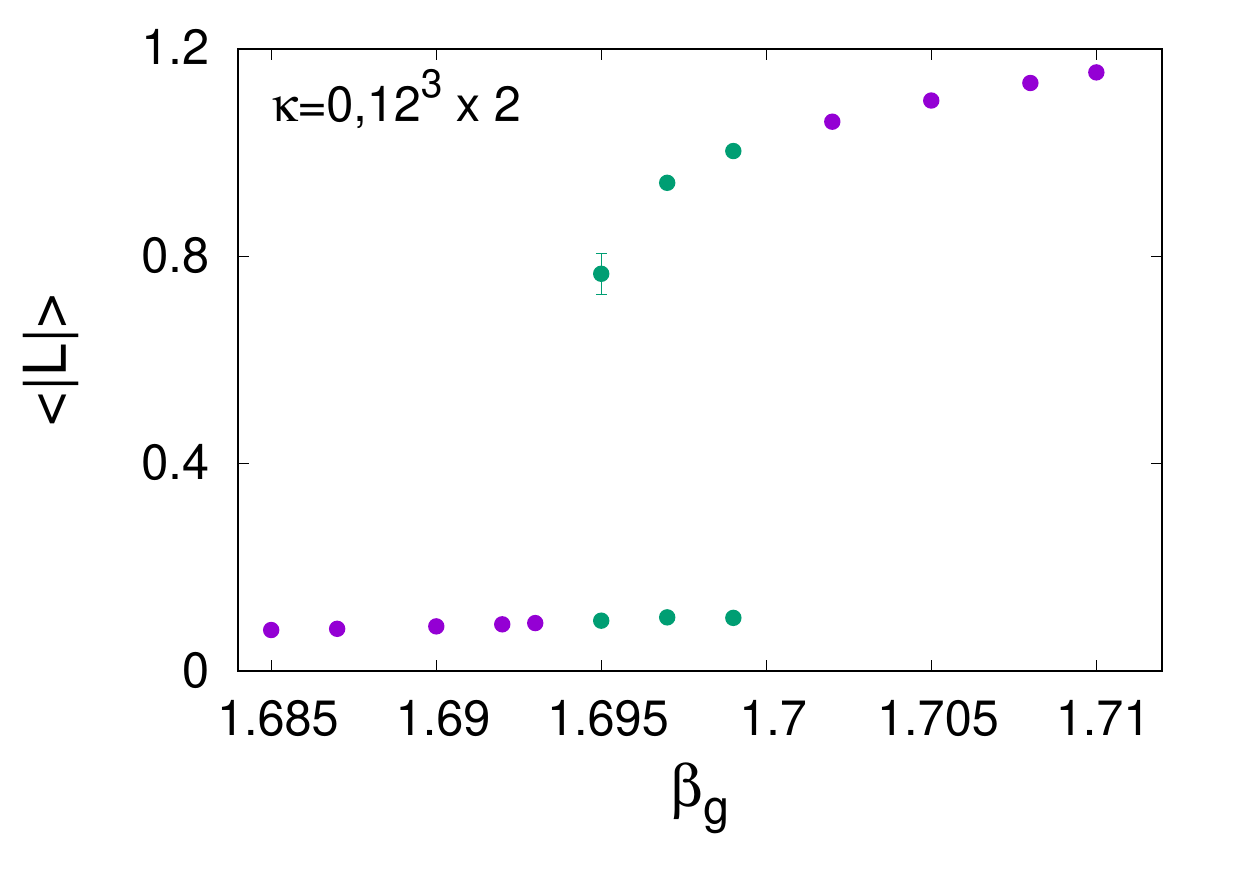}
    \caption{$\left<|L|\right>$ vs $\beta_g$ for $N_\tau=2$.}
    \label{pure_nt2}
  \end{minipage}
  \hfill
  \begin{minipage}[b]{0.48\textwidth}
    \includegraphics[width=\textwidth]{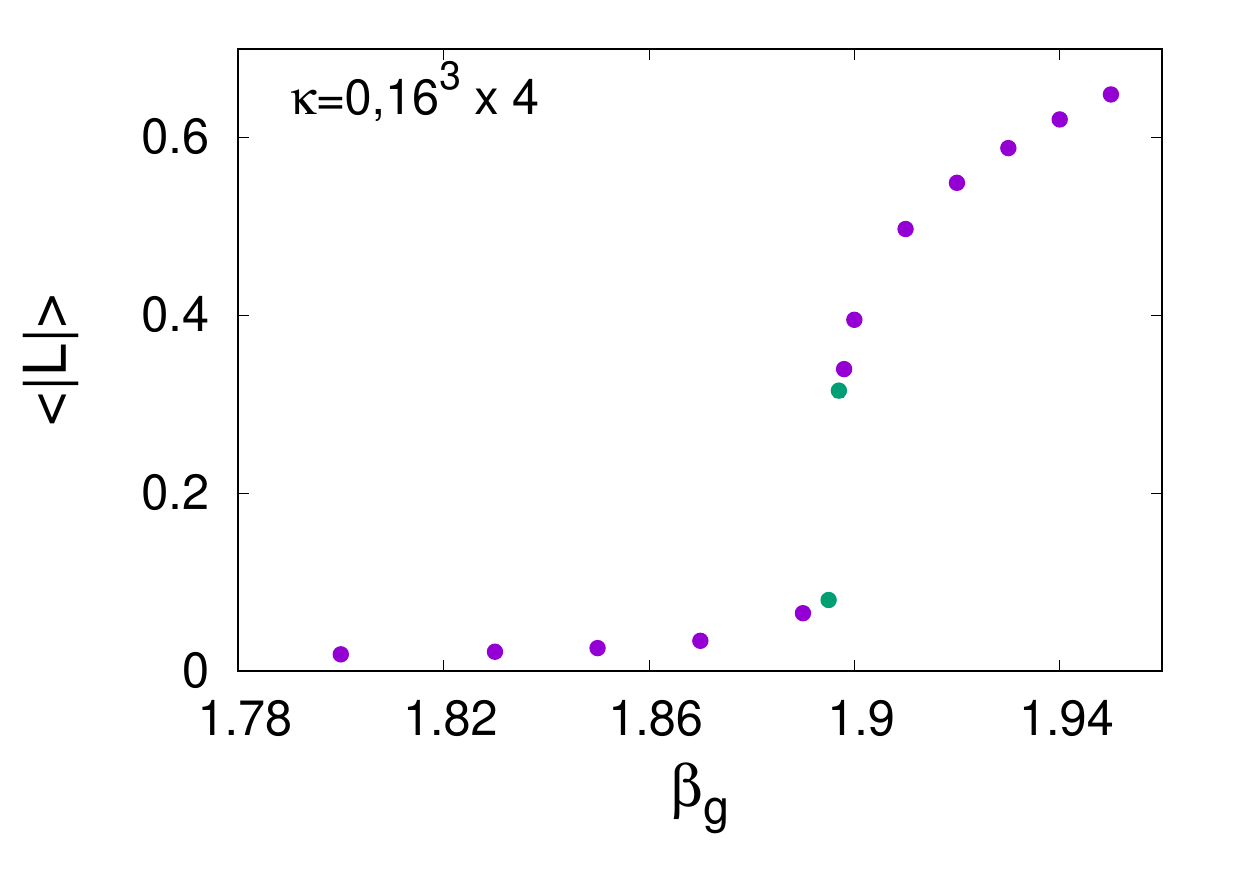}
    \caption{$\left<|L|\right>$ vs $\beta_g$ for $N_\tau=4$.}
    \label{pure_nt4}
  \end{minipage}
  \end{figure}
\begin{figure}[h]
	\centering
	\begin{minipage}[b]{0.48\textwidth}
		\includegraphics[width=\textwidth]{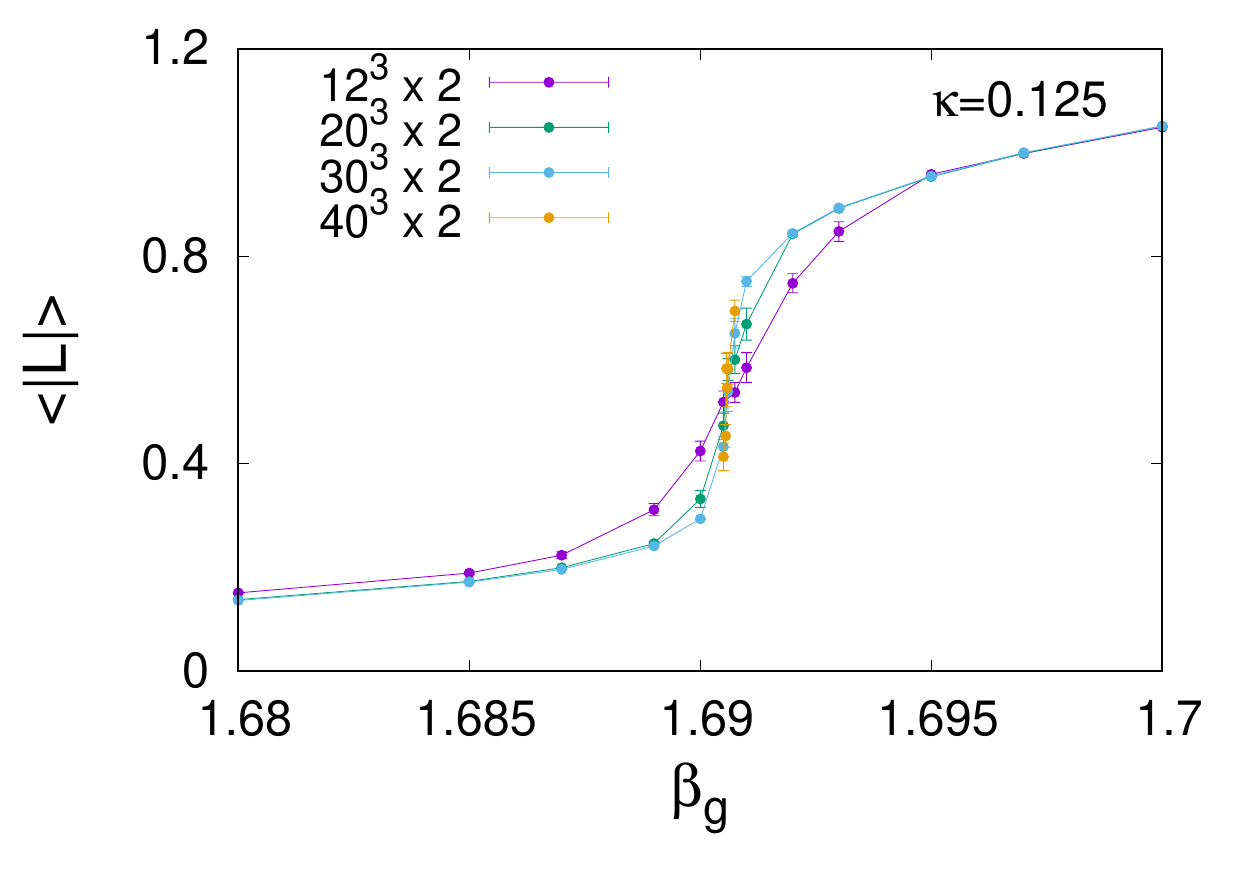}
		\caption{Polyakov loop vs $\beta_g$ for $\kappa=0.125$.}
		\label{su3h_nt2_m}
	\end{minipage}
	\hfill
	\begin{minipage}[b]{0.48\textwidth}
		\includegraphics[width=\textwidth]{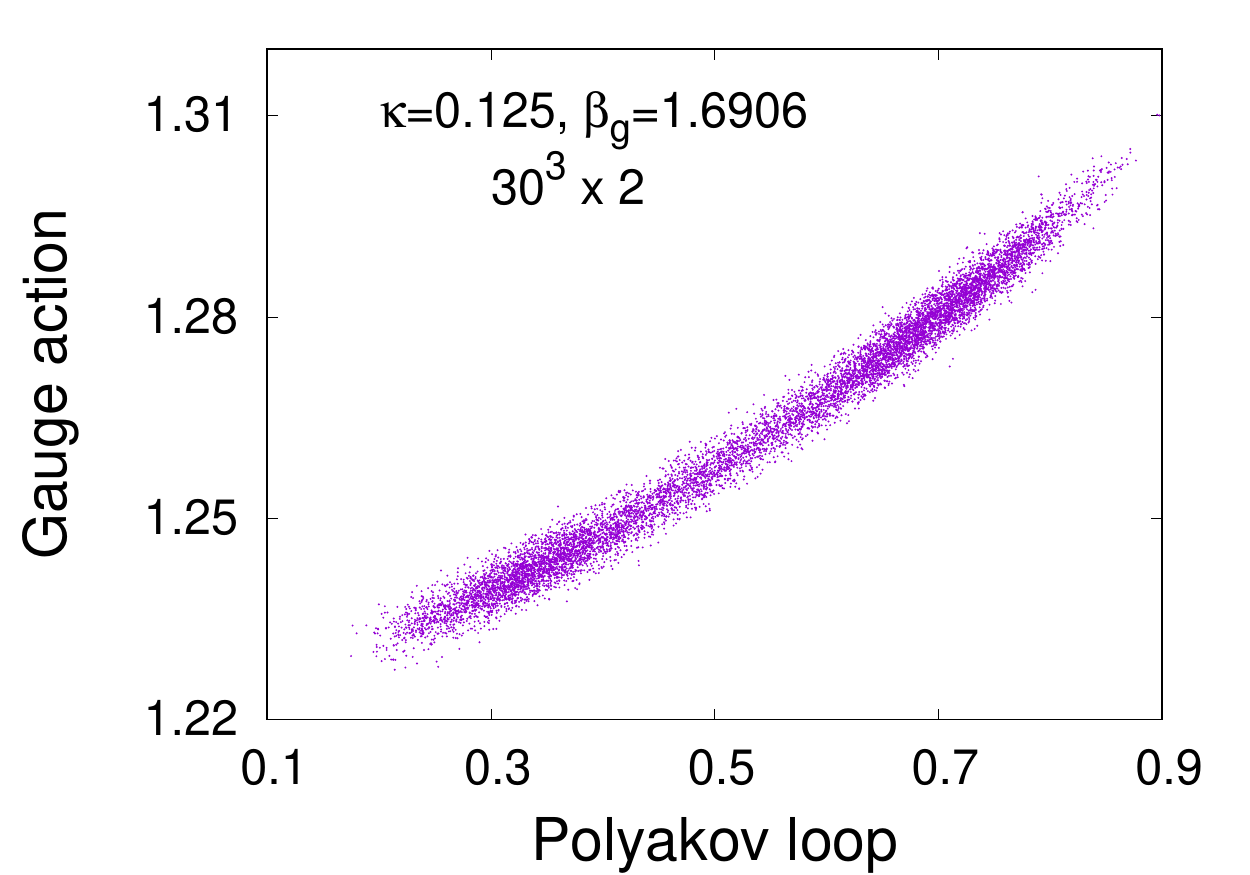}
		\caption{$S_g$ vs Polyakov loop at $\kappa=0.125$.}
		\label{su3h_nt2_s}
	\end{minipage}
\end{figure}
\begin{figure}[h]
	\centering
	\begin{minipage}[b]{0.48\textwidth}
		\includegraphics[width=\textwidth]{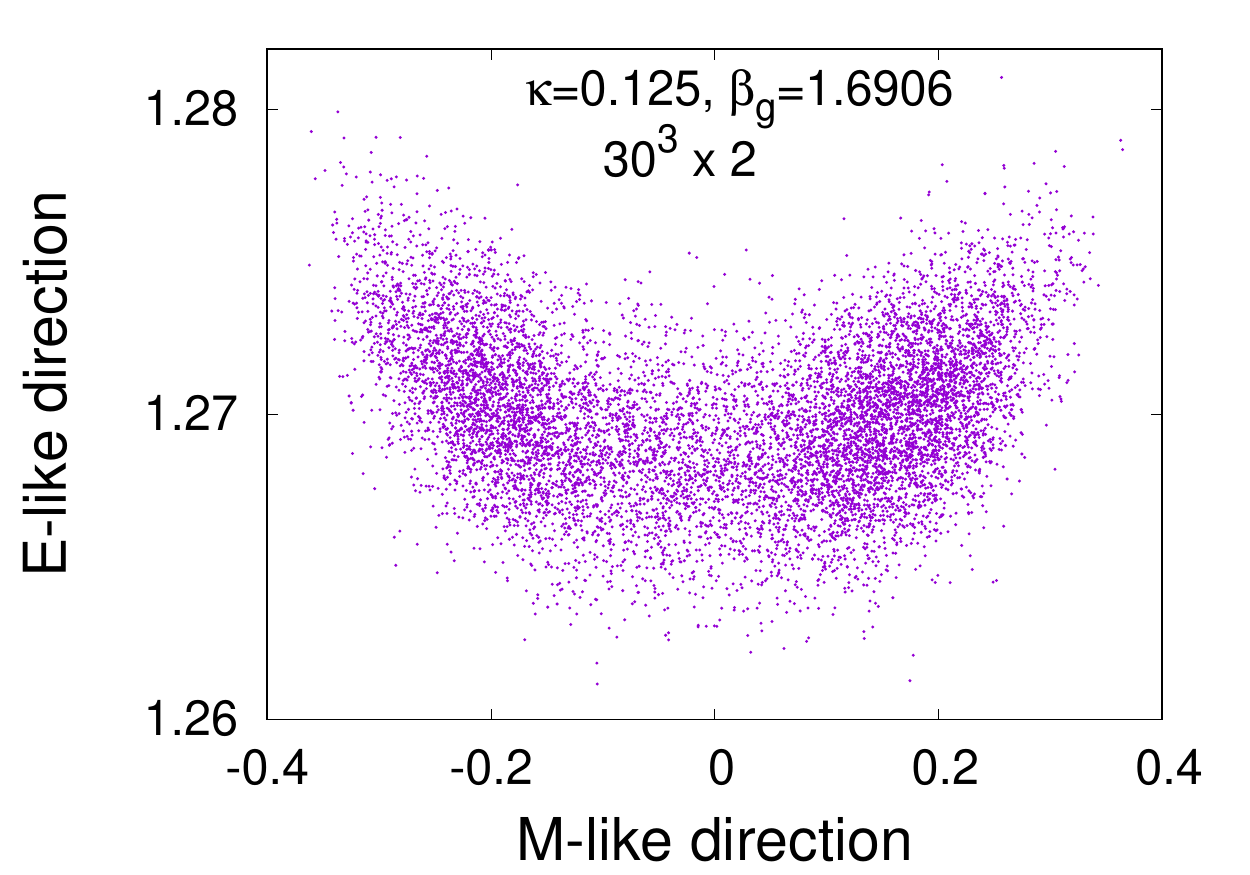}
		\caption{Gauge action vs Polyakov loop for $\kappa=0.125$.}
		\label{ga_nt2}
	\end{minipage}
	\hfill
	\begin{minipage}[b]{0.48\textwidth}
		\includegraphics[width=\textwidth]{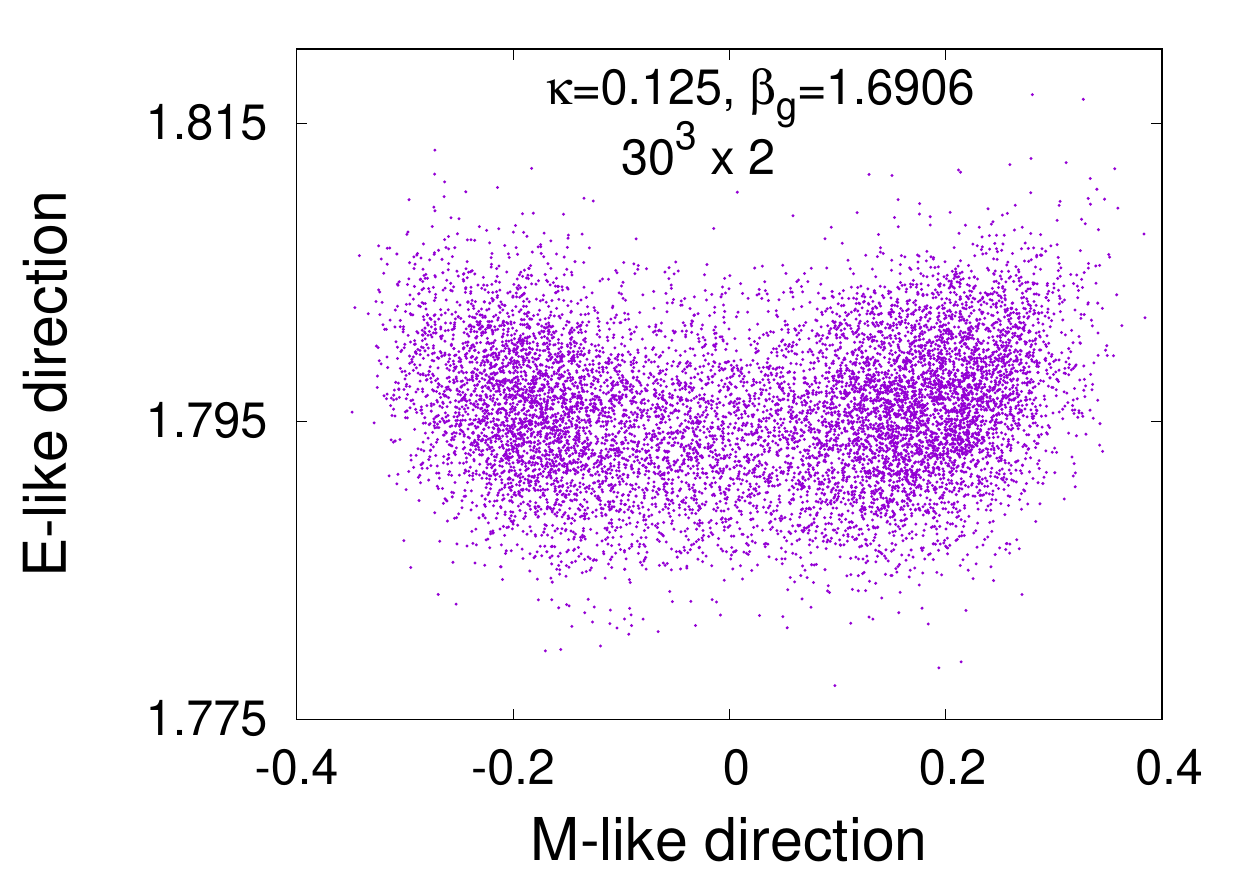}
		\caption{Total action vs Polyakov loop for $\kappa=0.125$.}
		\label{ta_nt2}
	\end{minipage}
\end{figure}

It is well established that, in pure $SU(3)$ gauge theory, the nature of the CD transition is first order ~\cite{Karsch:2001nf,Yaffe:1982qf,Celik:1983wz,Iwasaki:1991pc,Kennedy:1984dk,Gottlieb:1985ug,Fukugita:1985xc,Brown:1988qe,Kogut:1982rt,Fukugita:1986rr}. For pure $SU(3)$, in Fig.~\ref{plhst_nt2} we show $H(|L|)$ for $N_\tau=2$ at $\beta_g=1.698$. In Fig.~\ref{plhst_nt4}, the same is plotted for $N_\tau=4$ at $\beta_g=1.897$. Since the transition is first order, for $\beta_g$ values near the transition point, the histogram shows two peaks. The peak corresponding to the smaller (higher) value of $|L|$ corresponds to the confined (deconfined) phase.  In Figs.~\ref{pure_nt2} and \ref{pure_nt4}, the Polyakov loop average ($L$) vs $\beta_g$ is plotted for $N_\tau=2$ and $N_\tau=4$, respectively. 
There is a range in $\beta_g$ for which there are two values of $|L|$ for
a given $\beta_g$. These two points do not correspond to the partition function
average of the Polyakov loop but represent the locations of the peak
positions of $H(|L|)$. The peak location which is closer to zero
corresponds to the confined state and the other to the deconfined state.
Note that the positions of the two peaks do not change with the initial
condition or configuration and the number of statistics. On the other hand, the
partition function average of the Polyakov loop depends on the number of
statistics as the system tunnels between the confined and deconfined states
and vice versa. As in previous studies~\cite{Kennedy:1984dk,Gottlieb:1985ug,Fukugita:1985xc}, the transition region shifts to higher values with $N_\tau$.
 In the presence of Higgs with $\kappa=1/8$ (Fig.~\ref{su3h_nt2_m}) show $|L|$ vs $\beta_g$ for $N_\tau=2$. The spatial sizes considered are $N_s=12, 20, 30, 40$. $|L|$ is continuous in $\beta_g$ for all spatial sizes, but the variation of $|L|$ near the transition point is sharper for larger spatial sizes, clearly showing finite-size effects. Furthermore, the value of $|L|$ at the transition point is nonzero and nearly independent of $N_s$, which suggests that this CD transition may be an end point of a first-order transition. Such end points have been studied
both in condensed matter as well as in high-energy physics models~\cite{Rummukainen:1998as, Karsch:2000xv, Alonso:1993tv, Janke:1996qb, wilding1997simulation}. 
Conventionally, energy (E-like) and magnetization (M-like) observables are crucial to analyze the nature of a second-order phase transition. 
The E-like and M-like observables are expected to be uncorrelated. Following the methods employed in Ref.~\cite{Rummukainen:1998as},
we consider orthogonal linear combinations of the gauge action ($S_g$) and the Polyakov loop $|L|$ to obtain the E-like and M-like observables.
In Fig.~\ref{su3h_nt2_s}, we have plotted the distribution of $S_g$ vs $L$. The distribution clearly shows that $S_g$ and $|L|$ are correlated.
But the distribution of E-like and M-like, in Fig.~\ref{ga_nt2}, clearly shows that the partition function average of their correlation is vanishingly small.
More importantly, the shape of this distribution strongly resembles the corresponding distribution in the case of the 3D Ising model. This suggests
that the nature of CD transition for $N_\tau=2$ is likely the end point of a first-order phase transition. We also did the same analysis by substituting $S_g$ with the total action. The distributions of the resulting E-like and M-like observables are shown in Fig.~\ref{ta_nt2}. 

\begin{figure}[h]
	\centering
	\begin{minipage}[b]{0.48\textwidth}
		\includegraphics[width=\textwidth]{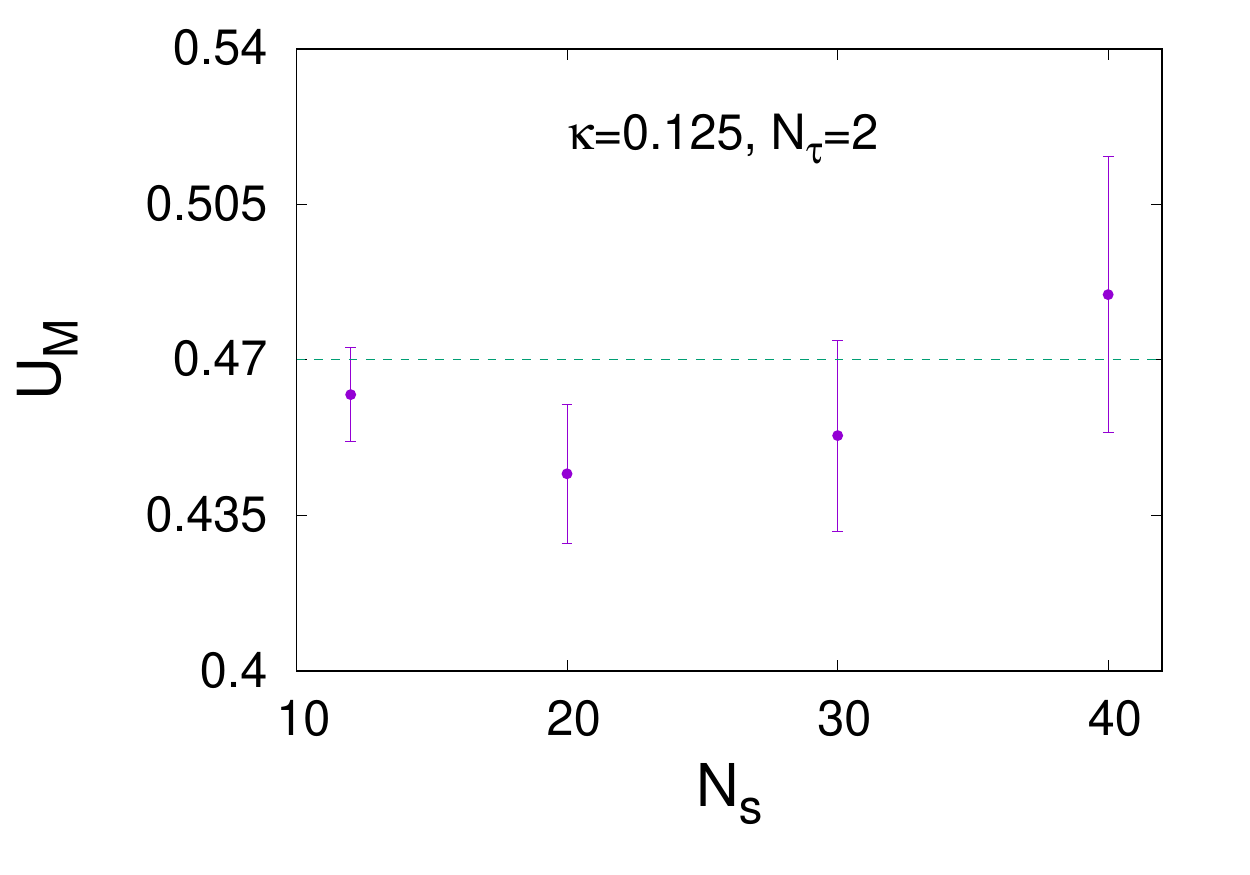}
		\caption{Binder cumulant vs spatial size ($N_s=12,20,30,40$).}
		\label{binder}
	\end{minipage}
	\hfill
	\begin{minipage}[b]{0.48\textwidth}
		\includegraphics[width=\textwidth]{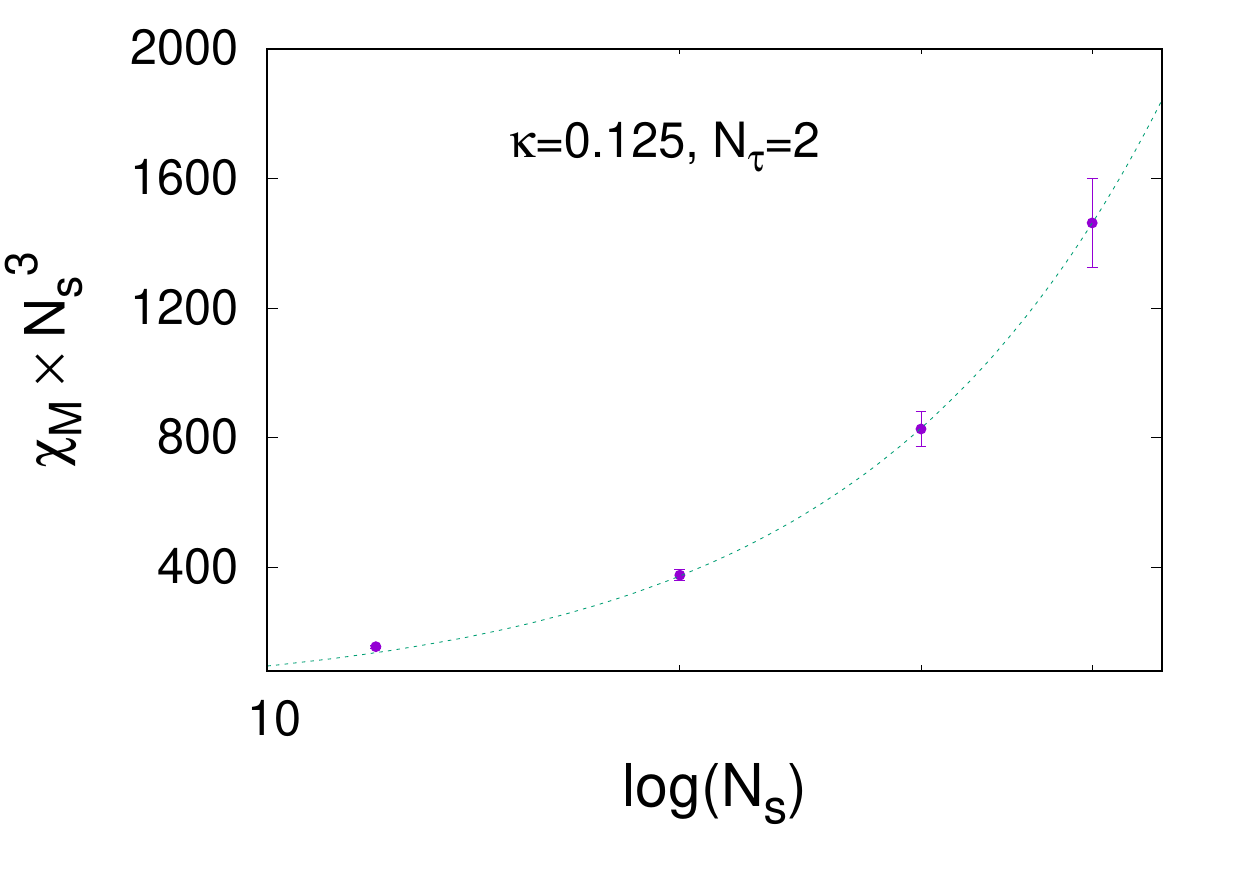}
		\caption{Susceptibility vs spatial size ($N_s=12,20,30,40$).}
		\label{sus}
	\end{minipage}
\end{figure}

To be more precise about the nature of the transition, the Binder cumulant ($U_M$) and the susceptibility ($\chi_M$) corresponding to the M-like observables are calculated near the critical point for different spatial lattice sizes. The value of the Binder cumulant near the critical point is within 0.44$-0.48$ for different $N_s$ as shown in Fig.~\ref{binder}; whereas for the 3D Ising universality class it is around 0.47~\cite{Biswal:2016xyq,Biswal:2015rul}. The critical exponent ``$\gamma/\nu$'' determined from the susceptibility of M-like observables ($\chi_M$) is shown in Fig.~\ref{sus} for $N_s=12,20,30,40$. The dotted line in Fig.~\ref{sus} corresponds to the fitted function giving the value of $\gamma/\nu \sim 1.973$. For 3D Ising universality class, the critical exponent $\gamma/\nu \sim 1.968$~\cite{Biswal:2016xyq,Rummukainen:1998as,Kanaya:1994qe}.  
So the observed $U_M$ and the critical indices are suggestive that it is an end point of a first-order phase transition.

\begin{figure}[h]
	\centering
	\begin{minipage}[b]{0.4\textwidth}
		\includegraphics[width=\textwidth]{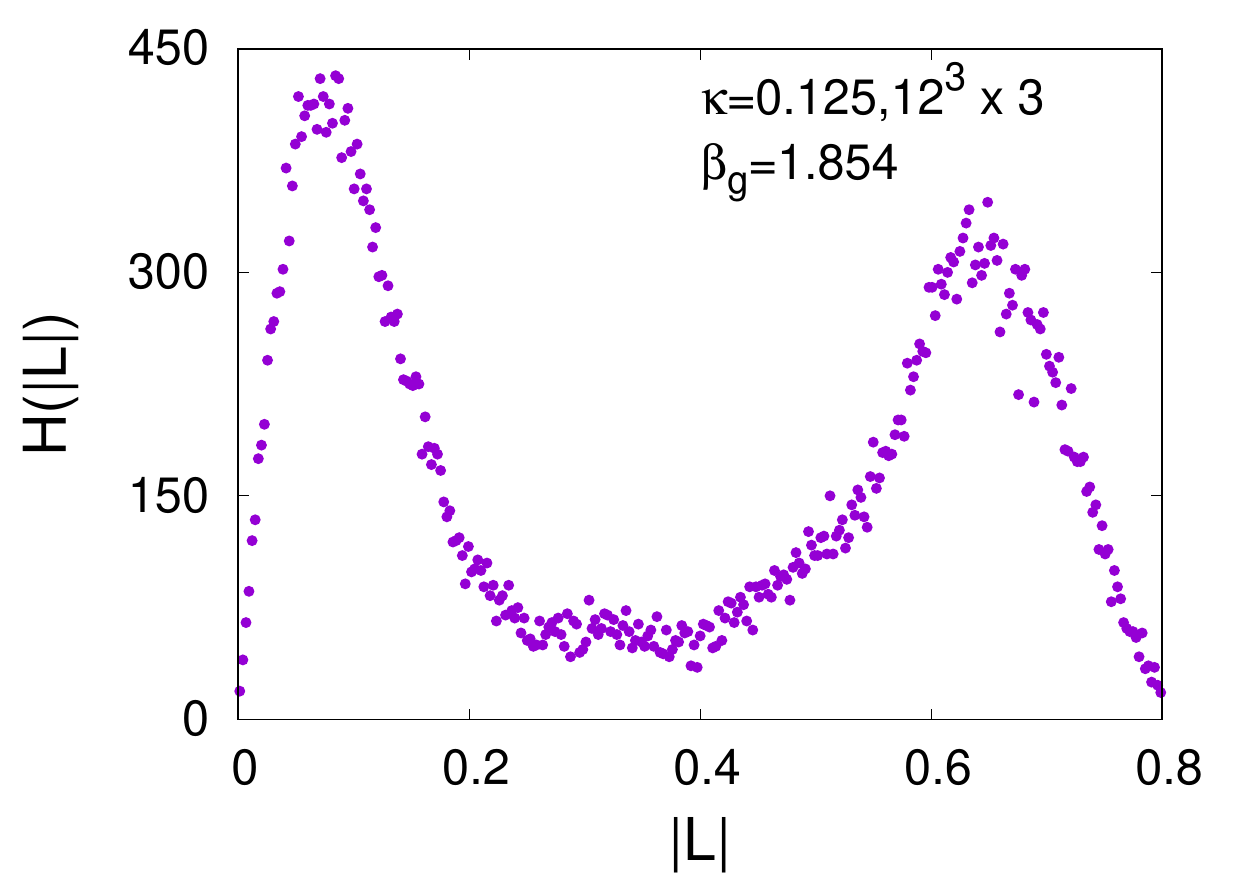}
		\caption{$H(|L|)$ for $N_\tau=3$ at $\beta_g=1.854$ and $\kappa=0.125$.}
		\label{plhst_nt3k8th}
	\end{minipage}
	\hfill
	\begin{minipage}[b]{0.4\textwidth}
		\includegraphics[width=\textwidth]{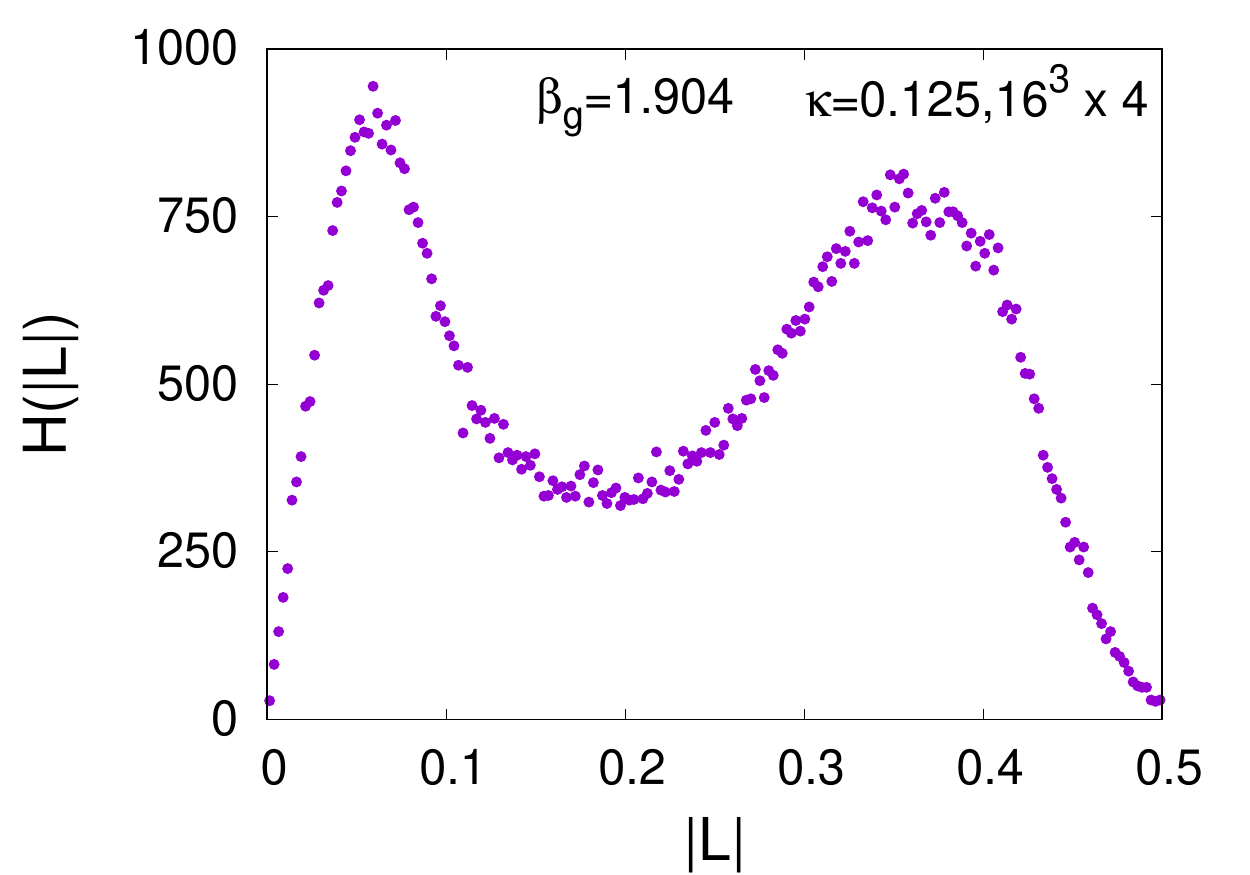}
		\caption{$H(|L|)$ for $N_\tau=4$ at $\beta_g=1.904$ and $\kappa=0.125$.}
		\label{plhst_nt4k8th}
	\end{minipage}
\end{figure}

\begin{figure}[h]
	\centering
	\begin{minipage}[b]{0.48\textwidth}
		\includegraphics[width=\textwidth]{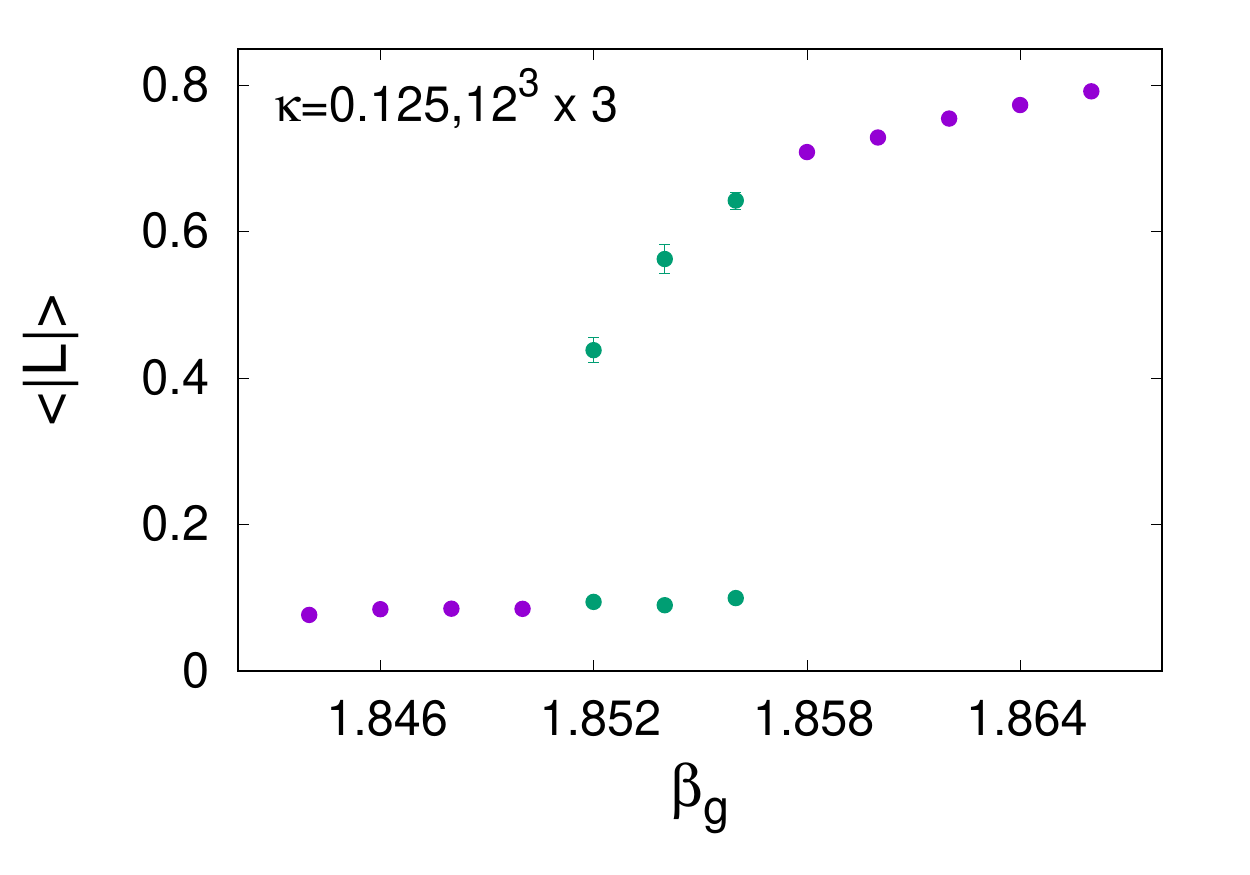}
		\caption{$\left<|L|\right>$ vs $\beta_g$ for $N_\tau=3$.}
		\label{su3h_nt3_m}
	\end{minipage}
	\hfill
	\begin{minipage}[b]{0.48\textwidth}
		\includegraphics[width=\textwidth]{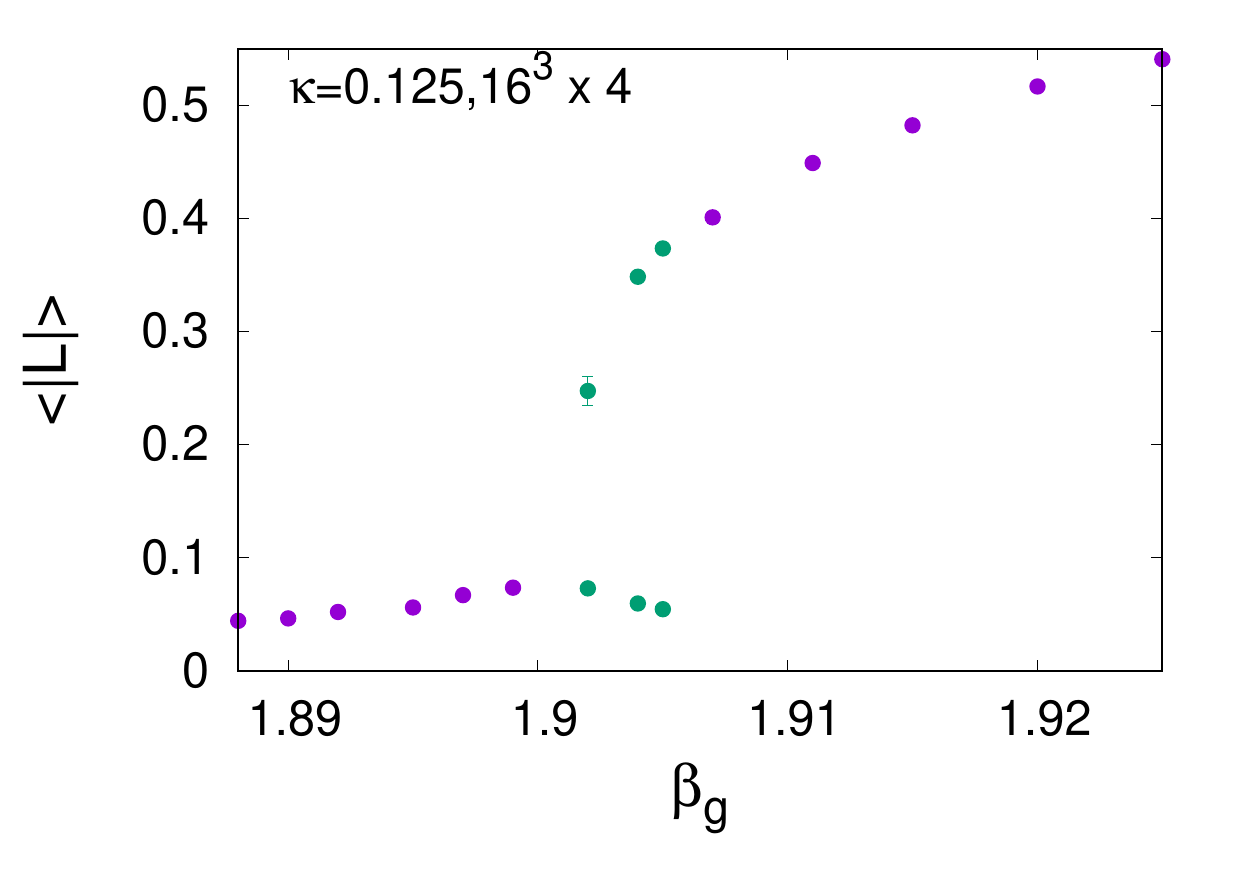}
		\caption{$\left<|L|\right>$ vs $\beta_g$ for $N_\tau=4$.}
		\label{su3h_nt4_m}
	\end{minipage}
\end{figure}

 Figures \ref{plhst_nt3k8th} and \ref{plhst_nt4k8th} show $H(|L|)$ for $N_\tau=3$ and $N_\tau=4$, at $\beta_g=1.854$ and $\beta_g=1.904$, respectively. $N_s=4N_\tau$. The results for $|L|$ vs $\beta_g$ are shown in Figs.~\ref{su3h_nt3_m} and \ref{su3h_nt4_m} for $N_\tau=3$ and $4$, respectively. These results suggest that the CD transition is first order. The finite-size scaling analysis for $N_\tau > 2$ shows that the results are independent of lattice sizes.
The results from $N_\tau=2$ to $N_\tau=4$ show that the nature of CD transition changes with $N_\tau$. For higher $N_\tau$ the CD transition continues to be first order. Since higher $N_\tau$ corresponds to a smaller cutoff, these results suggest that the CD transition will be first order in the
 continuum limit.  
 \newpage
 \subsection{$Z_3$ symmetry vs $N_\tau$} 
  In this section, we present observables which are sensitive to the $Z_3$ symmetry, i.e., the distribution of the Polyakov loop in the complex plane, the average of the gauge-Higgs interaction $S_K$, and the gauge action $S_g$. When there is $Z_3$ symmetry, the distribution should be invariant, when
the transformation $L \to z L$ is made. Furthermore, in the deconfined phase, the difference of $S_K$ between different
 $Z_3$ states should vanish. Here, $Z_3$ states refer to states for which the Polyakov loop phase ($\theta$) is $0,2\pi/3$, or $4\pi/3$. 
 
    \begin{figure}[h]
  \centering
      \begin{minipage}[b]{0.48\textwidth}
    \includegraphics[width=\textwidth]{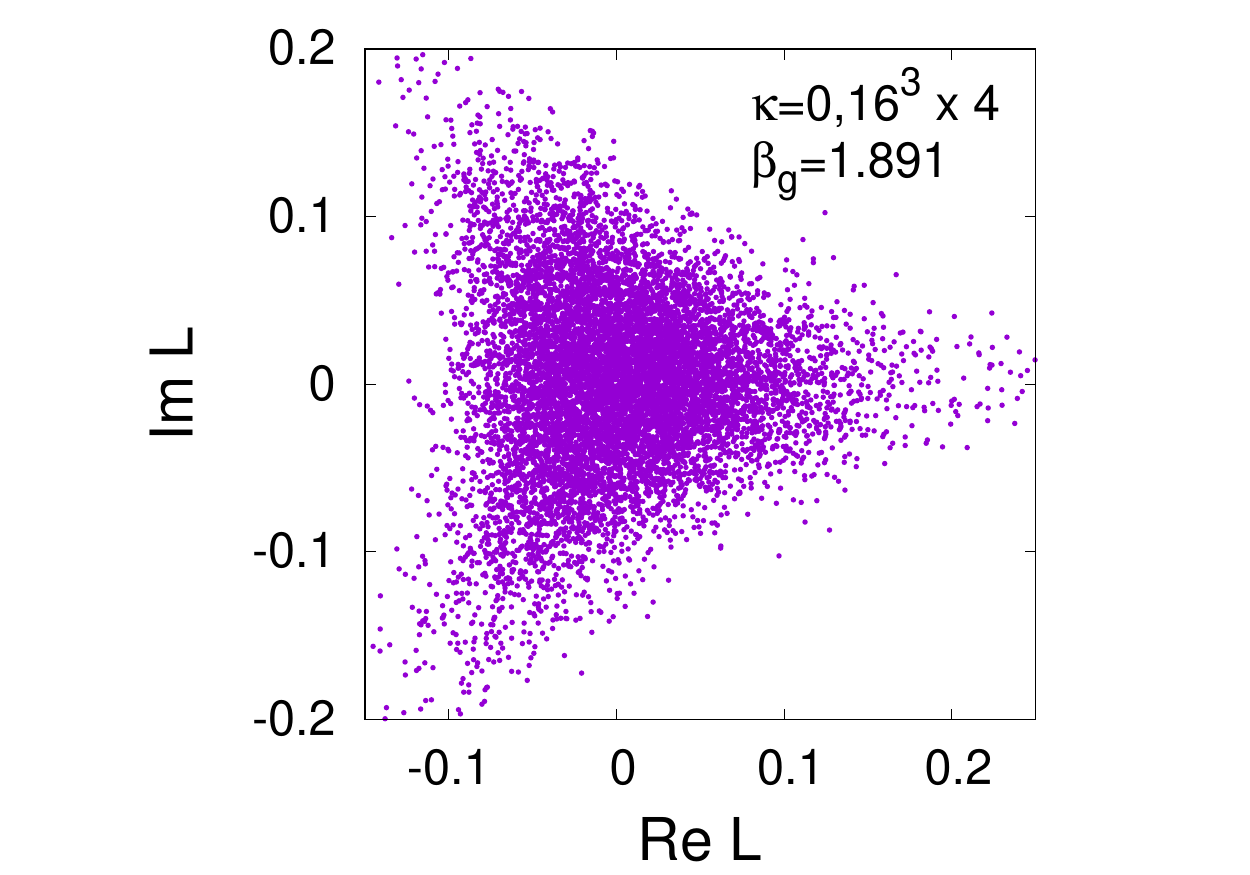}
    \caption{Distribution of $L$ in the confined phase for $N_\tau=4$.}
    \label{pdist_nt4_conf}
  \end{minipage}
  \hfill
      \begin{minipage}[b]{0.48\textwidth}
    \includegraphics[width=\textwidth]{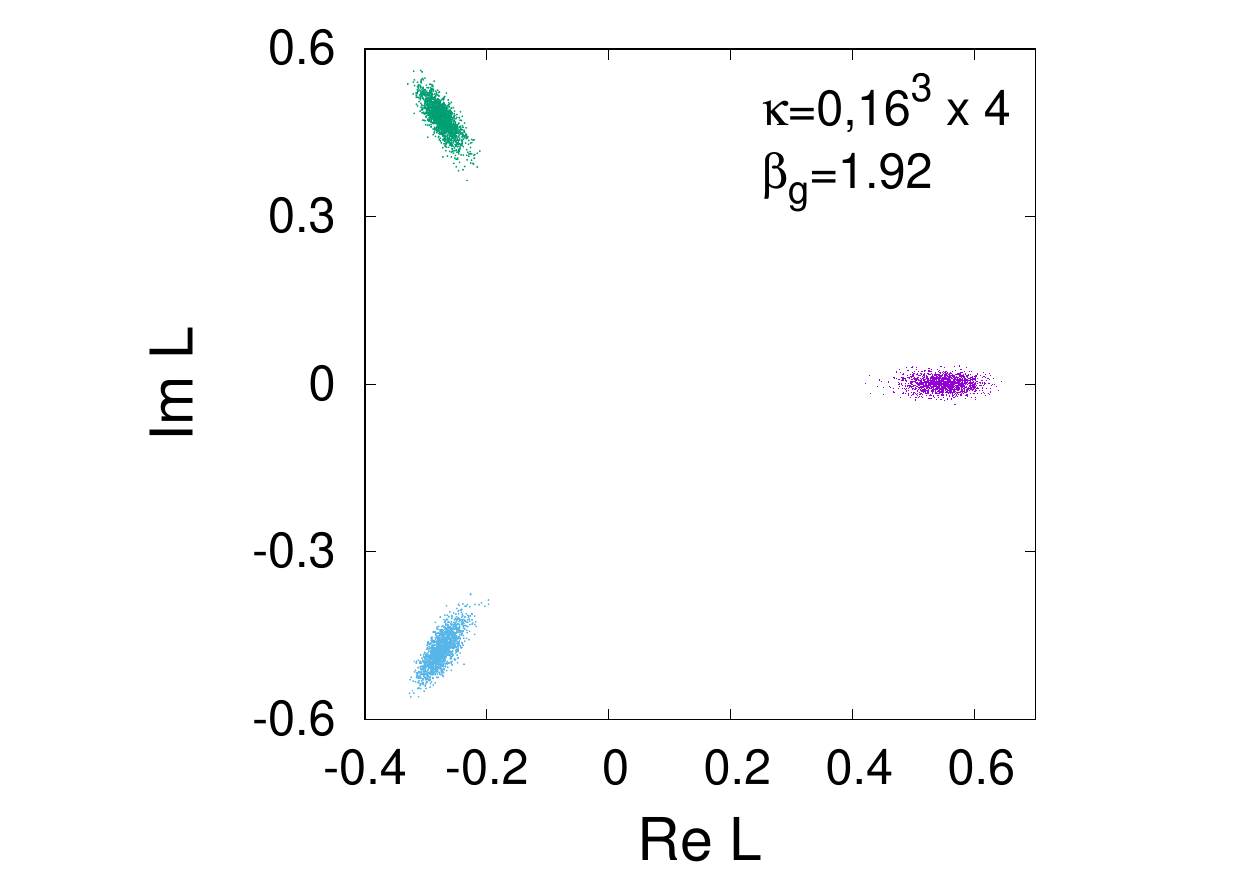}
    \caption{Distribution of $L$ in the deconfined phase for $N_\tau=4$.}
    \label{pdist_nt4_dconf}
      \end{minipage}
   \end{figure}  
 \par
\begin{figure}[H]
  \centering
    \begin{minipage}[b]{0.465\textwidth}
      \includegraphics[width=\textwidth]{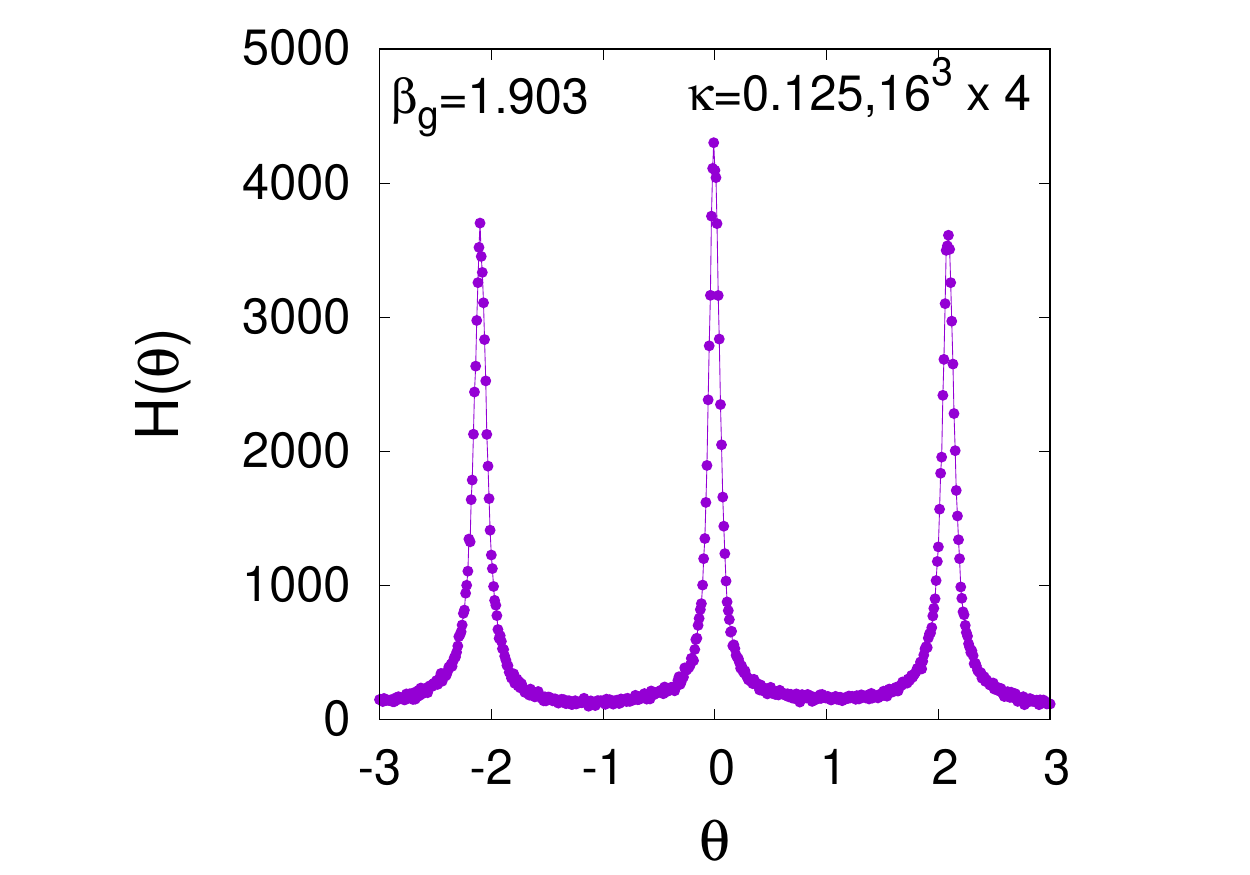}
 \caption{Distribution of phase of the Polyakov loop.}
 \label{hth_nt4}
  \end{minipage}
  \hfill
  \begin{minipage}[b]{0.48\textwidth}
        \includegraphics[width=\textwidth]{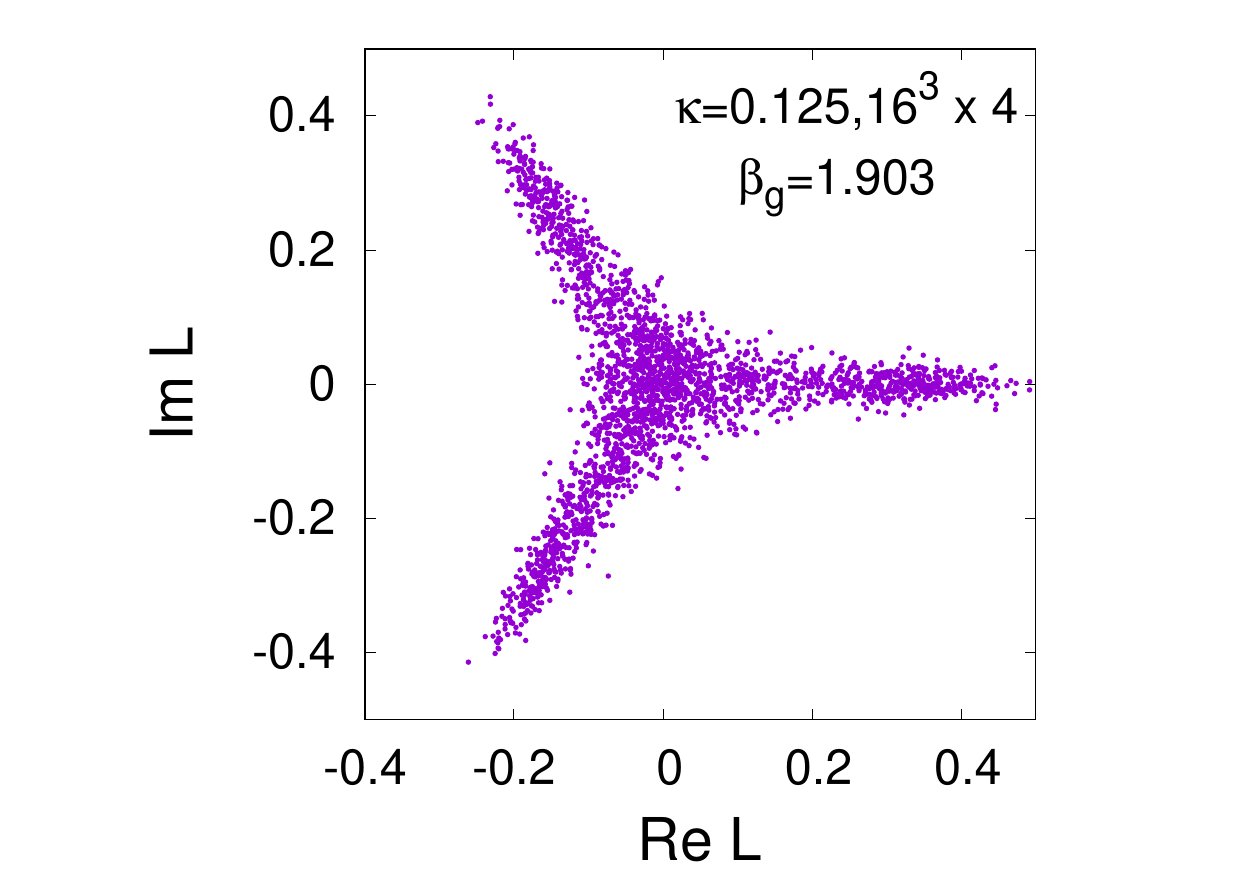}
    \caption{$L$ on the complex plane for $16^3 \times 4$ lattice.}
 \label{distpl_nt4}
  \end{minipage}
 \end{figure}

 The distributions of $L$ for pure $SU(3)$ are shown in Figs.~\ref{pdist_nt4_conf} and \ref{pdist_nt4_dconf} at $\beta_g=1.891$
and $\beta_g=1.92$, respectively. The distribution in Fig.~\ref{pdist_nt4_conf} corresponds to the confined phase and in Fig.~\ref{pdist_nt4_dconf} corresponds to the deconfined phase. There is $Z_3$ symmetry in both these distributions. In the deconfined phase, $\beta_g>\beta_{gc}$, the symmetry is spontaneously broken, which leads to $Z_3$ states. The three patches in Fig.~\ref{pdist_nt4_dconf} correspond to the three $Z_3$ states. Note that all three states, for $\beta_g$ away from $\beta_{gc}$, cannot be sampled in a single MC run, as the tunnelling rate between them is very small. To sample different $Z_3$ states, we consider MC runs with different initial conditions. Though the Polyakov loop values differ, they have the same free energy.
\begin{figure}[h]
 \centering
  \includegraphics[width=0.6\textwidth]{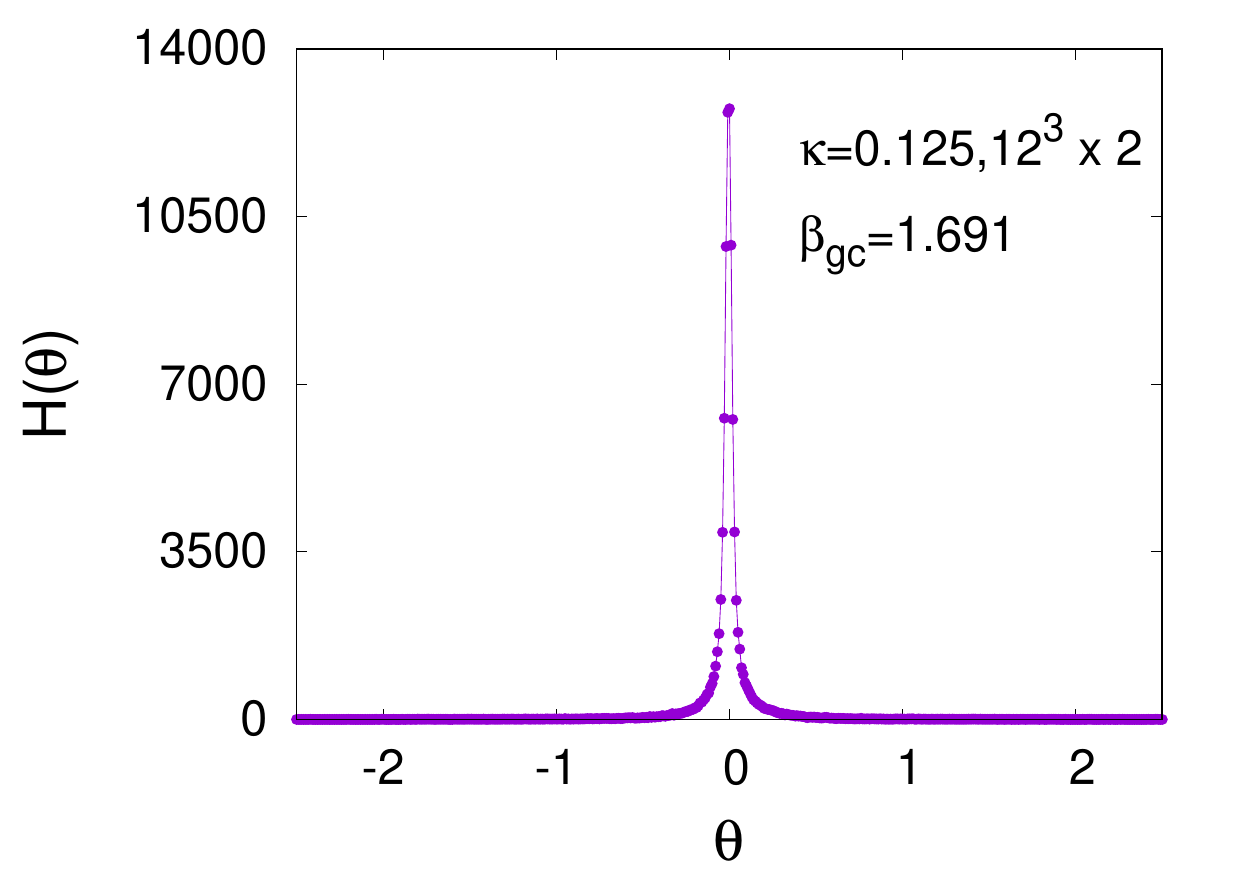}
 \caption{Distribution of phase of the Polyakov loop for $\left<|L|\right>=0.584874$.}
 \label{hth_nt2}
 \end{figure}
In the presence of Higgs, in Fig.~\ref{hth_nt2}, the distribution $H(\theta)$  vs $\theta$ is plotted at $\beta_g=1.691$ for $N_\tau=2$. $H(\theta)$ vs $\theta$ has only one peak at $\theta=0$. The $Z_3$ symmetry is clearly broken, as there are no peaks corresponding to $\theta=2\pi/3,4\pi/3$. For $N_\tau=2$, the distribution
of $L$ exhibits $Z_3$ symmetry in both the confined and deconfined phases. In the presence of Higgs, for $N_\tau=2$, even though there is 
explicit breaking, the $\theta=2\pi/3,4\pi/3$ states develop deep in the deconfinement phase. For $N_\tau=4$, Fig.~\ref{hth_nt4} shows $H(\theta)$
close to the critical point. There is a slight $Z_3$ asymmetry in $H(\theta)$, though peaks corresponding to $\theta=2\pi/3,4\pi/3$ are almost
comparable to that at $\theta=0$. In Fig.~\ref{distpl_nt4}, the measured values of $L$,  for the same $\beta_g$, are plotted in the complex 
plane. The distribution of the scattered point is almost $Z_3$ symmetric.

 \par
\begin{figure}[H]
  \centering
    \begin{minipage}[b]{0.465\textwidth}
      \includegraphics[width=\textwidth]{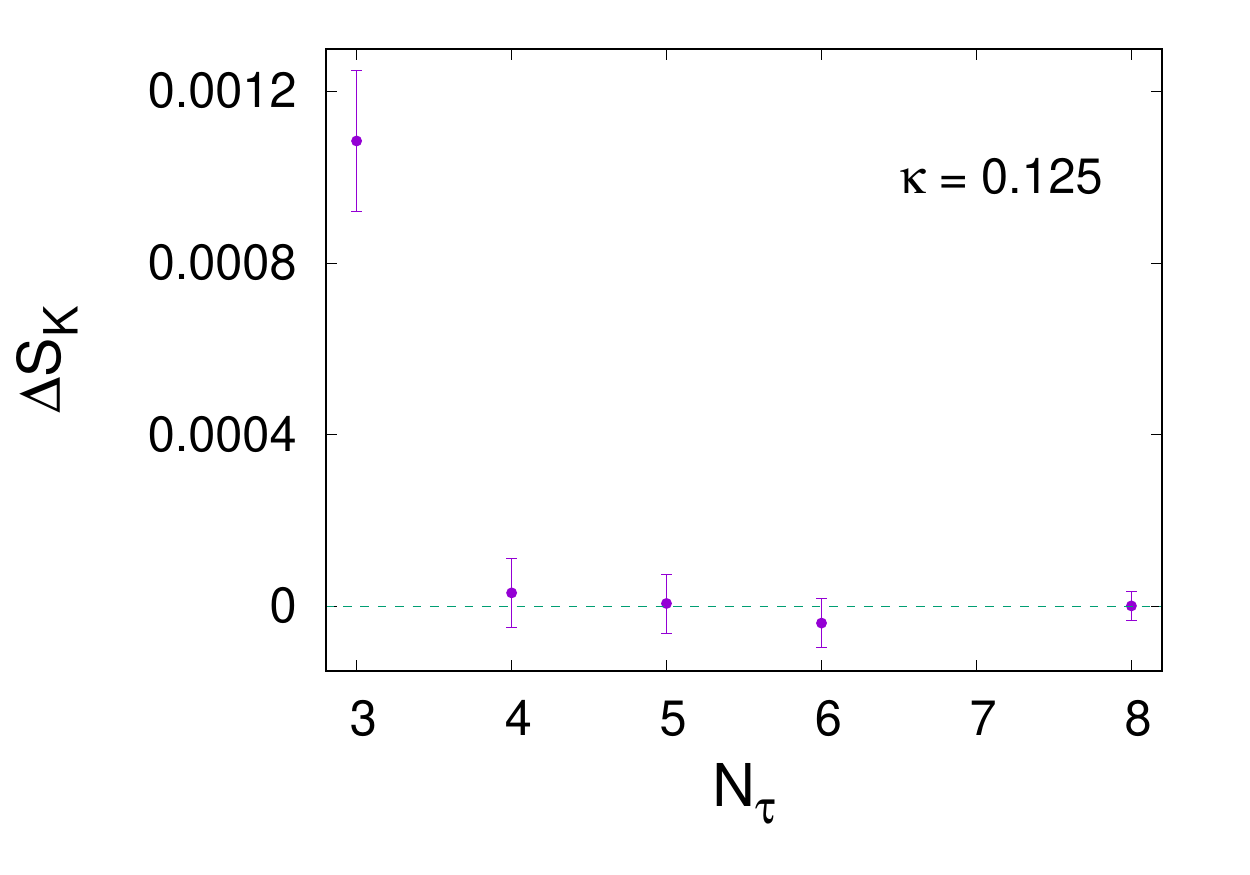}
 \caption{Difference of $S_K$ between $\theta=0$ and $2\pi/3$ in the deconfinement phase.}
 \label{0_vs_z3_sk}
  \end{minipage}
  \hfill
  \begin{minipage}[b]{0.48\textwidth}
        \includegraphics[width=\textwidth]{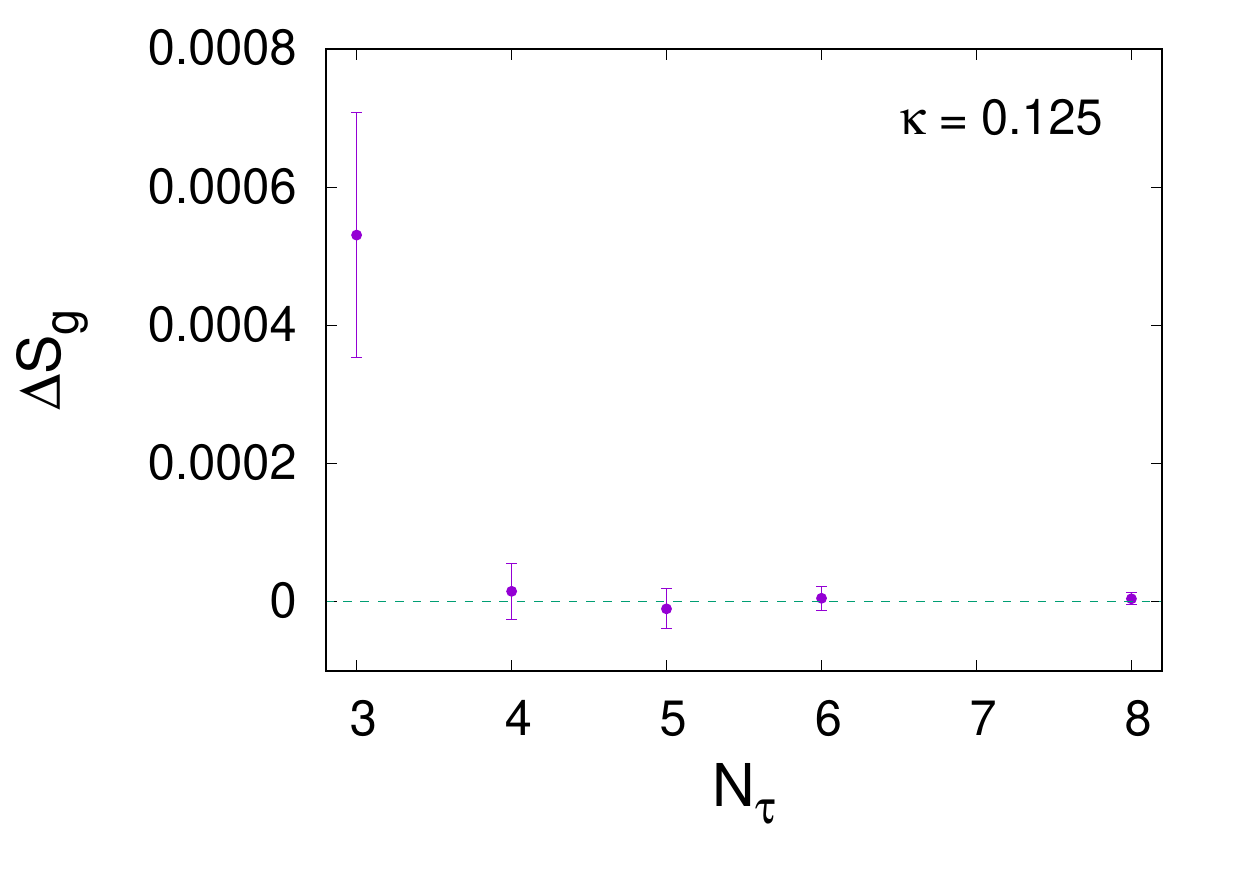}
    \caption{Difference of $S_g$ between $\theta=0$ and $2\pi/3$ in the deconfinement phase.}
 \label{0_vs_z3_sg}
  \end{minipage}
 \end{figure}

 These results suggest that for larger $N_\tau$ the explicit breaking of $Z_3$ near the transition point significantly decreases. To compare the physical properties of the $Z_3$ states, we compute $\Delta S_K=S_K(\theta=0)-S_K(\theta=2\pi/3)$ and $\Delta S_g=S_g(\theta=0)-S_g(\theta=2\pi/3)$ for different $N_\tau$ at the same physical temperature. To fix the temperature, the $\beta_g$ values for different $N_\tau$ are obtained from the one-loop beta-function~\cite{Damgaard:1986jg}. $\Delta S_K$ and $\Delta S_g$ vs $N_\tau$ are plotted in Figs.~\ref{0_vs_z3_sk} and \ref{0_vs_z3_sg}, respectively. The results show that the differences between $\theta=0$ and $\theta=2\pi/3$ states exponentially decrease. The free energy difference between these states can be calculated by integrating $\Delta S_K(\kappa)$ over $\kappa = \{0,1/8\}$, in other words, over $m_H = $($\infty,0\}$. Since $\Delta S_K(\kappa=0) = 0$, the integrand is vanishingly small over the integration range. As a consequence, the difference will also be vanishingly small in the continuum limit.  

Note that the physical mass of the Higgs will be nonzero even though
the bare mass is vanishingly small. This mass is due to the fluctuations
resulting from gauge-Higgs interaction. On the lattice, physical Higgs 
mass differs from the continuum limit for small $N_\tau$. For the $N_\tau$ values considered in 
our study, the critical temperature computed from the one-loop beta 
function decreases with $N_\tau$. The decrease can be well approximated 
by an exponential with a nonzero limiting value, which is indicative of the
cutoff effects decreasing with $N_\tau$. Our results for the critical 
temperature suggest that the difference in cutoff effects between 
$N_\tau=6$ and $N_\tau=8$ is much less compared to that between $N_\tau=2$ 
and $N_\tau=4$. Similarly, We expect that the difference in physical Higgs 
mass between $N_\tau=6$ and $N_\tau=8$ will be smaller compared to that 
between $N_\tau=2$ and $N_\tau=4$, and it will approach a nonzero 
value in the continuum limit. Since the explicit breaking of $Z_3$ 
decreases monotonically with $N_\tau$ and becomes vanishingly small 
(Figs.~\ref{0_vs_z3_sk} and \ref{0_vs_z3_sg}), within statistical errors, 
for $N_\tau=6$ and $8$, we expect it to remain so even for larger $N_\tau$.

 \par
\begin{figure}[h]
 \centering
  \includegraphics[width=0.48\textwidth]{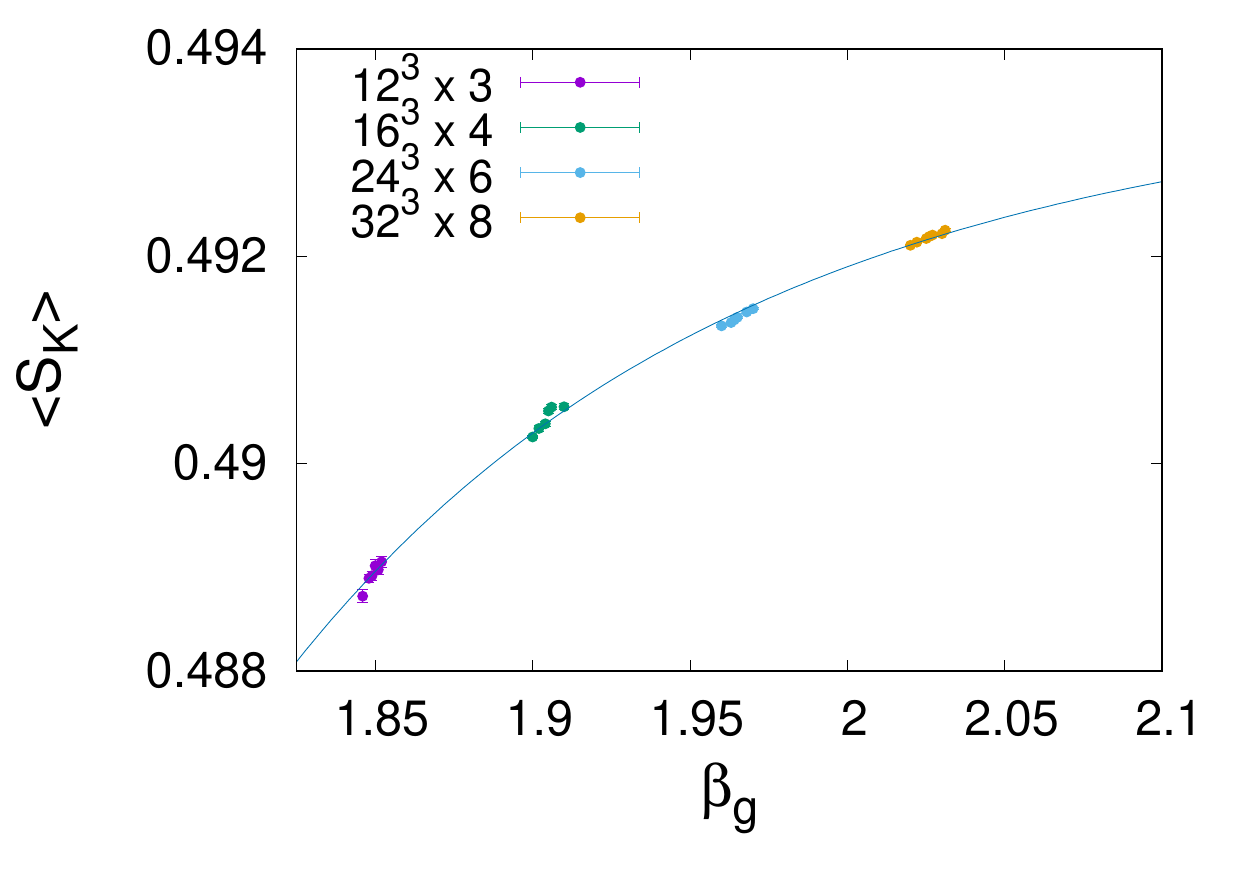}
 \caption{$S_K$ for different $N_\tau$ near $\beta_{gc}$.}
 \label{sk_vs_nt}
 \end{figure}

 The above results suggest that the explicit breaking $Z_3$ symmetry will be vanishingly small in the continuum limit. To test whether the decrease in $Z_3$ explicit breaking with $N_\tau$ is due to a decrease in the interaction between the gauge and Higgs fields with $N_\tau$, we compare the gauge Higgs interaction term ($S_K$) in Eq.~(\ref{lac}). A weaker interaction with decreasing $N_\tau$ should lead to a decrease in $S_K$. Our results, in Fig.~\ref{sk_vs_nt}, show that $S_K$ increases monotonically with $N_\tau$. Note that $S_K$ in physical units will also increase, as the lattice spacing decreases with $N_\tau$. The estimation of this increase requires the critical $\beta_{g}$ vs $N_\tau$. Because of the coexistence of the confined and deconfined states near the transition point, it is difficult to find the critical value of $\beta_g$ accurately.

 \section{Conclusions}
 
 We have studied the CD transition and $Z_3$ symmetry in $SU(3)$-Higgs theory for vanishing bare Higgs mass and quartic coupling. Most of the MC simulations that have been done are around the CD transition point. The Monte Carlo results show that the nature of the CD transition and the explicit breaking of $Z_3$ vary with $N_\tau$. For $N_\tau=2$, analysis of the simulation results using conventional methods suggests that the CD transition is an end point of a first-order phase transition. The distribution of the Polyakov loop breaks $Z_3$ symmetry, with no peaks corresponding to the $Z_3$ symmetry.
For $N_\tau=3$, the transition is first order. The distribution of the Polyakov loop near the transition point does have peaks corresponding to all the $Z_3$ sectors. However, the peak heights are not the same, breaking the $Z_3$ symmetry.  This suggests that the explicit breaking is there but small compared to that of the $N_\tau=2$ case. 
 
 The explicit breaking for $N_\tau=4$ is similar compared to $N_\tau=3$. The distributions of the Polyakov loop show partial $Z_3$ symmetry with a smaller difference in the peak heights of $Z_3$ sectors compared to $N_\tau=3$. This pattern that the CD transition is first order and decrease in the explicit breaking continues for higher $N_\tau$ in our simulations. To make a quantitative assessment of explicit breaking, we
compute the difference of the gauge-Higgs interaction as well as that of the pure gauge part of the action, between different $Z_3$ states. Our results show that both observables' differences decrease exponentially with $N_\tau$. The vanishing difference in the large $N_\tau$ limit will lead to the same free energy for all the $Z_3$ states. These results suggest that the CD transition is first order and the explicit breaking of $Z_3$ is vanishingly small in the continuum limit. In the presence of higher-order terms in the Higgs potential, our results
will be still valid as long as the system is in the Higgs symmetric part of
the phase diagram, except near second-order Higgs phase transition points. We
mention here that there are no studies on the $Z_N$ symmetry when a   
second-order Higgs transition point or line is approached from the Higgs
symmetric side of the phase diagram. At a second-order transition point or line, 
the fluctuations of the physical observables involving the Higgs field will 
diverge, including the interaction term. It is not clear that in this case the 
$Z_N$ symmetry will be realized.
 
The action we consider in this study does not possess $Z_3$ symmetry, but the partition function averages turn out to be $Z_3$ symmetric.  We believe that this may be due to the dominance of the $Z_3$ symmetric entropy, over the Boltzmann factor in the continuum limit. The vanishing of the explicit breaking for vanishing Higgs mass and quartic coupling should also hold for the nonzero Higgs mass case. In the future, we plan to study the implications of nonzero $\lambda$.  We mention here that the perturbative calculations show that deep inside the deconfinement $Z_3$ is explicitly broken. It is possible that the realization of $Z_3$ is limited to the region close to the transition point. It will be interesting to explore $Z_3$ for large $\beta_g$ values and compare them with perturbative results.

\vspace{1cm}


\centerline{\bf  REFERENCES}\vskip -20pt
\begin{thebibliography}{99}

\bibitem{Nakamura:1984uz}
A.~Nakamura,
Phys. Lett. B \textbf{149}, 391 (1984)
doi:10.1016/0370-2693(84)90430-1



\bibitem{Fukugita:1986rr}
M.~Fukugita and A.~Ukawa,
Phys. Rev. Lett. \textbf{57}, 503 (1986)
doi:10.1103/PhysRevLett.57.503



\bibitem{Kogut:1982rt}
J.~B.~Kogut, M.~Stone, H.~W.~Wyld, W.~R.~Gibbs, J.~Shigemitsu, S.~H.~Shenker and D.~K.~Sinclair,
Phys. Rev. Lett. \textbf{50}, 393 (1983)
doi:10.1103/PhysRevLett.50.393


\bibitem{Karsch:2001nf}
F.~Karsch, E.~Laermann and C.~Schmidt,
Phys. Lett. B \textbf{520}, 41-49 (2001)
doi:10.1016/S0370-2693(01)01114-5
[arXiv:hep-lat/0107020 [hep-lat]].



\bibitem{Polyakov:1978vu}
A.~M.~Polyakov,
Phys. Lett. B \textbf{72}, 477-480 (1978)
doi:10.1016/0370-2693(78)90737-2



\bibitem{Susskind:1979up}
L.~Susskind,
Phys. Rev. D \textbf{20}, 2610-2618 (1979)
doi:10.1103/PhysRevD.20.2610


\bibitem{Green:1982hu}
F.~Green,
Nucl. Phys. B \textbf{215}, 83-108 (1983)
doi:10.1016/0550-3213(83)90268-7


\bibitem{Pisarski:1982cn}
R.~D.~Pisarski and O.~Alvarez,
Phys. Rev. D \textbf{26}, 3735 (1982)
doi:10.1103/PhysRevD.26.3735



\bibitem{Gavai:1985nz}
R.~V.~Gavai and F.~Karsch,
Nucl. Phys. B \textbf{261}, 273-284 (1985)
doi:10.1016/0550-3213(85)90575-9


\bibitem{McLerran:1981pb}
L.~D.~McLerran and B.~Svetitsky,
Phys. Rev. D \textbf{24}, 450 (1981)
doi:10.1103/PhysRevD.24.450



\bibitem{McLerran:1980pk}
L.~D.~McLerran and B.~Svetitsky,
Phys. Lett. B \textbf{98}, 195 (1981)
doi:10.1016/0370-2693(81)90986-2




\bibitem{Kuti:1980gh}
J.~Kuti, J.~Polonyi and K.~Szlachanyi,
Phys. Lett. B \textbf{98}, 199 (1981)
doi:10.1016/0370-2693(81)90987-4



\bibitem{Svetitsky:1982gs}
B.~Svetitsky and L.~G.~Yaffe,
Nucl. Phys. B \textbf{210}, 423-447 (1982)
doi:10.1016/0550-3213(82)90172-9



\bibitem{Weiss:1980rj}
N.~Weiss,
Phys. Rev. D \textbf{24}, 475 (1981)
doi:10.1103/PhysRevD.24.475




\bibitem{Svetitsky:1985ye}
B.~Svetitsky,
Phys. Rept. \textbf{132}, 1-53 (1986)
doi:10.1016/0370-1573(86)90014-1





\bibitem{Yaffe:1982qf}
L.~G.~Yaffe and B.~Svetitsky,
Phys. Rev. D \textbf{26}, 963 (1982)
doi:10.1103/PhysRevD.26.963


\bibitem{Celik:1983wz}
T.~Celik, J.~Engels and H.~Satz,
Phys. Lett. B \textbf{125}, 411-414 (1983)
doi:10.1016/0370-2693(83)91314-X


\bibitem{Weiss:1981ev}
N.~Weiss,
Phys. Rev. D \textbf{25}, 2667 (1982)
doi:10.1103/PhysRevD.25.2667


\bibitem{Belyaev:1991np}
V.~M.~Belyaev, I.~I.~Kogan, G.~W.~Semenoff and N.~Weiss,
Phys. Lett. B \textbf{277}, 331-336 (1992)
doi:10.1016/0370-2693(92)90754-R



\bibitem{Green:1983sd}
F.~Green and F.~Karsch,
Nucl. Phys. B \textbf{238}, 297-306 (1984)
doi:10.1016/0550-3213(84)90452-8


\bibitem{Biswal:2016xyq}
M.~Biswal, M.~Deka, S.~Digal and P.~S.~Saumia,
Phys. Rev. D \textbf{96}, no.1, 014503 (2017)
doi:10.1103/PhysRevD.96.014503
[arXiv:1610.08265 [hep-lat]].



\bibitem{Biswal:2015rul}
M.~Biswal, S.~Digal and P.~S.~Saumia,
Nucl. Phys. B \textbf{910}, 30-39 (2016)
doi:10.1016/j.nuclphysb.2016.06.025
[arXiv:1511.08295 [hep-lat]].

\bibitem{Hasenfratz:1983ce}
P.~Hasenfratz, F.~Karsch and I.~O.~Stamatescu,
Phys. Lett. B \textbf{133}, 221-226 (1983)
doi:10.1016/0370-2693(83)90565-8

\bibitem{Heller:1984eq}
U.~M.~Heller and F.~Karsch,
Nucl. Phys. B \textbf{258}, 29-45 (1985)
doi:10.1016/0550-3213(85)90601-7


\bibitem{Kogut:1985xd}
J.~B.~Kogut, J.~Polonyi, H.~W.~Wyld and D.~K.~Sinclair,
Phys. Rev. D \textbf{31}, 3307 (1985)
doi:10.1103/PhysRevD.31.3307


\bibitem{Heller:1985wc}
U.~M.~Heller,
Phys. Lett. B \textbf{163}, 203-206 (1985)
doi:10.1016/0370-2693(85)90221-7




\bibitem{Ignatius:1991nk}
J.~Ignatius, K.~Kajantie and K.~Rummukainen,
Phys. Rev. Lett. \textbf{68}, 737-740 (1992)
doi:10.1103/PhysRevLett.68.737



\bibitem{Dixit:1991et}
V.~Dixit and M.~C.~Ogilvie,
Phys. Lett. B \textbf{269}, 353-356 (1991)
doi:10.1016/0370-2693(91)90183-Q



\bibitem{Biswal:2019xju}
M.~Biswal, S.~Digal and P.~S.~Saumia,
Phys. Rev. D \textbf{102}, no.7, 074020 (2020)
doi:10.1103/PhysRevD.102.074020
[arXiv:1907.07981 [hep-ph]].



\bibitem{Fradkin:1978dv}
E.~H.~Fradkin and S.~H.~Shenker,
Phys. Rev. D \textbf{19}, 3682-3697 (1979)
doi:10.1103/PhysRevD.19.3682



\bibitem{Deka:2010bc}
M.~Deka, S.~Digal and A.~P.~Mishra,
Phys. Rev. D \textbf{85}, 114505 (2012)
doi:10.1103/PhysRevD.85.114505
[arXiv:1009.0739 [hep-lat]].



\bibitem{Biswal:2018ilm}
M.~Biswal,``$Z _N$ Symmetry and confinement-deconfinement transition in $SU(N)$+Higgs theory,''
http://www.hbni.ac.in/phdthesis/phys/PHYS10201204004.pdf



\bibitem{Guo:2018scp}
Y.~Guo and Q.~Du,
JHEP \textbf{05}, 042 (2019)
doi:10.1007/JHEP05(2019)042
[arXiv:1810.13090 [hep-ph]].



\bibitem{Gross:1980br}
D.~J.~Gross, R.~D.~Pisarski and L.~G.~Yaffe,
Rev. Mod. Phys. \textbf{53}, 43 (1981)
doi:10.1103/RevModPhys.53.43



\bibitem{Satz:1985js}
H.~Satz,
Phys. Lett. B \textbf{157}, 65-69 (1985)
doi:10.1016/0370-2693(85)91213-4



\bibitem{Damgaard:1986jg}
P.~H.~Damgaard and U.~M.~Heller,
Phys. Lett. B \textbf{171}, 442-448 (1986)
doi:10.1016/0370-2693(86)91436-X




\bibitem{Biswal:2021mhp}
M.~Biswal, S.~Digal, V.~Mamale and S.~Shaikh,
Int. J. Mod. Phys. A \textbf{37}, no.09, 2250047 (2022)
doi:10.1142/S0217751X22500476
[arXiv:2102.12935 [hep-lat]].



\bibitem{Biswal:2021fde}
M.~Biswal, S.~Digal, V.~Mamale and S.~Shaikh,
Mod. Phys. Lett. A \textbf{36}, no.30, 2150218 (2021)
doi:10.1142/S0217732321502187
[arXiv:2102.11091 [hep-lat]].




\bibitem{Kajantie:1995kf}
K.~Kajantie, M.~Laine, K.~Rummukainen and M.~E.~Shaposhnikov,
Nucl. Phys. B \textbf{466}, 189-258 (1996)
doi:10.1016/0550-3213(96)00052-1
[arXiv:hep-lat/9510020 [hep-lat]].



\bibitem{Cabibbo:1982zn}
N.~Cabibbo and E.~Marinari,
Phys. Lett. B \textbf{119}, 387-390 (1982)
doi:10.1016/0370-2693(82)90696-7



\bibitem{Kennedy:1985nu}
A.~D.~Kennedy and B.~J.~Pendleton,
Phys. Lett. B \textbf{156}, 393-399 (1985)
doi:10.1016/0370-2693(85)91632-6



\bibitem{Iwasaki:1991pc}
Y.~Iwasaki, K.~Kanaya, T.~Yoshie, T.~Hoshino, T.~Shirakawa, Y.~Oyanagi, S.~Ichii and T.~Kawai,
Phys. Rev. Lett. \textbf{67}, 3343-3346 (1991)
doi:10.1103/PhysRevLett.67.3343



\bibitem{Kennedy:1984dk}
A.~D.~Kennedy, J.~Kuti, S.~Meyer and B.~J.~Pendleton,
Phys. Rev. Lett. \textbf{54}, 87 (1985)
doi:10.1103/PhysRevLett.54.87



\bibitem{Fukugita:1985xc}
M.~Fukugita, T.~Kaneko and A.~Ukawa,
Phys. Lett. B \textbf{154}, 185-189 (1985)
doi:10.1016/0370-2693(85)90581-7




\bibitem{Gottlieb:1985ug}
S.~A.~Gottlieb, J.~Kuti, D.~Toussaint, A.~D.~Kennedy, S.~Meyer, B.~J.~Pendleton and R.~L.~Sugar,
Phys. Rev. Lett. \textbf{55}, 1958 (1985)
doi:10.1103/PhysRevLett.55.1958

\bibitem{Brown:1988qe}
F.~R.~Brown, N.~H.~Christ, Y.~F.~Deng, M.~S.~Gao and T.~J.~Woch,
Phys. Rev. Lett. \textbf{61}, 2058 (1988)
doi:10.1103/PhysRevLett.61.2058

\bibitem{Rummukainen:1998as}
K.~Rummukainen, M.~Tsypin, K.~Kajantie, M.~Laine and M.~E.~Shaposhnikov,
Nucl. Phys. B \textbf{532}, 283-314 (1998)
doi:10.1016/S0550-3213(98)00494-5
[arXiv:hep-lat/9805013 [hep-lat]].



\bibitem{Karsch:2000xv}
F.~Karsch and S.~Stickan,
Phys. Lett. B \textbf{488}, 319-325 (2000)
doi:10.1016/S0370-2693(00)00902-3
[arXiv:hep-lat/0007019 [hep-lat]].


\bibitem{Alonso:1993tv}
J.~L.~Alonso, V.~Azcoiti, I.~Campos, J.~C.~Ciria, A.~Cruz, D.~Iniguez, F.~Lesmes, C.~Piedrafita, A.~Rivero and A.~Tarancon, \textit{et al.}
Nucl. Phys. B \textbf{405}, 574-592 (1993)
doi:10.1016/0550-3213(93)90560-C
[arXiv:hep-lat/9210014 [hep-lat]].


\bibitem{Janke:1996qb}
W.~Janke and R.~Villanova,
Nucl. Phys. B \textbf{489}, 679-696 (1997)
doi:10.1016/S0550-3213(96)00710-9
[arXiv:hep-lat/9612008 [hep-lat]].



\bibitem{wilding1997simulation}
Wilding, Nigel B,
Journal of Physics: Condensed Matter \textbf{9}, 585 (1997)
doi:10.1088/0953-8984/9/3/002
[arXiv:cond-mat/9610133 ].


\bibitem{Kanaya:1994qe}
K.~Kanaya and S.~Kaya,
Phys. Rev. D \textbf{51}, 2404-2410 (1995)
doi:10.1103/PhysRevD.51.2404
[arXiv:hep-lat/9409001 [hep-lat]].













\end{thebibliography}
\end{document}